\PassOptionsToPackage{table,xcdraw}{xcolor}
\documentclass[acmsmall,screen,nonacm]{acmart}
\AtBeginDocument{%
  }

\setcopyright{none}

\usepackage[most]{tcolorbox}
\usepackage{multirow}
\usepackage{mdframed}
\usepackage{pifont}
\usepackage{enumitem}
\newcommand{\unmarkedfootnote}[1]{%
  \begingroup
  \renewcommand{\thefootnote}{}%
  \begin{NoHyper}\footnotetext{#1}\end{NoHyper}%
  \addtocounter{footnote}{-1}%
  \endgroup
}
\usepackage{amsmath}

\usepackage{svg}
\svgpath{{./}{./figures/}}
\tcbset{textmarker/.style={%
        enhanced, parbox=false,boxrule=0mm,boxsep=0mm,arc=0mm,
        outer arc=0mm,left=1mm,right=1mm,top=7pt,bottom=7pt,
        toptitle=1mm,bottomtitle=1mm,oversize}}
\newtcolorbox{importantBox}{textmarker,
    borderline west={2pt}{0pt}{red},
    colback=red!10!white}

\newcommand{\bcircle}[1]{\ding{\numexpr181 + #1}}

\newcommand{\answer}[2]{
  \begin{tcolorbox}[
    left=3mm,
    right=3mm,
    left skip=4pt, 
    colback=gray!10,
    colframe=gray!80,
    boxrule=0.5pt,
    leftrule=4pt, 
    top=1mm,
    before skip=0mm,
    after skip=0mm
    ]
    \textbf{Insight #1:}
    #2
  \end{tcolorbox}
}

\def\figlabel#1{\label{fig:#1}\label{p:#1}}

\usepackage{xcolor}
\usepackage{amsmath}
\usepackage[capitalize,noabbrev]{cleveref}
\usepackage{float}

\newcommand{\ourapproach}{\textit{CodeAnchor}}
\newcommand{\rev}[1]{{\color{black}#1}}
\usepackage{pifont}

\usepackage{array} 
\usepackage{url}
\begin{document}

\title[How Much Static Structure Do Code Agents Need?]{How Much Static Structure Do Code Agents Need? A Study of Deterministic Anchoring}








\author{Zhihao Lin}
\affiliation{%
  \institution{Beihang University}
  \city{Beijing}
  \country{China}
}
\email{mathieulin@buaa.edu.cn}

\author{Mingyi Zhou}
\authornotemark[1]
\affiliation{%
  \institution{Beihang University}
  \city{Beijing}
  \country{China}
}
\email{zhoumingyi@buaa.edu.cn}

\author{Yizhuo Yang}
\affiliation{%
  \institution{Beihang University}
  \city{Beijing}
  \country{China}
}
\email{yangyizhuo@buaa.edu.cn}

\author{Li Li}
\authornote{Corresponding author.}
\affiliation{%
  \institution{Beihang University}
  \city{Beijing}
  \country{China}
}
\email{lilicoding@ieee.org}

\renewcommand{\shortauthors}{Lin et al.}
\begin{abstract}
LLM-based code agents navigate repositories through keyword search but miss the structural relationships, such as call graphs, inheritance hierarchies, and configuration dependencies, that define how software actually works. This makes agent navigation stochastic and difficult to reproduce across runs.
We investigate whether lightweight static analysis can provide \emph{deterministic anchors} for these agents: stable structural facts injected as plain-text comments that constrain probabilistic exploration and make navigation more predictable. Starting from a strong baseline, Codex from OpenAI, we systematically inject varying granularities of structural annotations and measure their effects on localization, trajectory behavior, and run-to-run stability.
Our study identifies what we call the \textbf{deterministic anchoring effect}: static structure helps less by making agents ``smarter'' and more by making their navigation \emph{disciplined and reproducible}. Three observations support this finding: \bcircle{1}~\textbf{Anchoring works}: lightweight call/inheritance topology improves function-level localization (+2.2pp Func@5) and shortens trajectories ($-$1.6 interaction rounds); \bcircle{2}~\textbf{Anchoring is scale-sensitive}: the optimal granularity and directionality depend on repository characteristics, where denser semantics show diminishing returns and hub-heavy projects benefit from inverse-only links that expose ``who-calls-me'' without forward edges; \bcircle{3}~\textbf{Anchoring stabilizes}: tags raise link-following rate from 0.15--0.18 to 0.21--0.24, roughly halve run-to-run variance, and improve single-run reliability (Pass@1 +3.4\,pp) on medium-scale repositories, at the cost of roughly 10\% more input tokens.
These observations suggest practical guidelines: default to lightweight topology on medium projects, prune forward edges in large repositories, and reserve dense tags for implicit-dependency cases. Code and traces are available in our repository: \url{https://github.com/mathieu0905/Code-Anchor}.
\end{abstract}

\keywords{Code agents, static analysis, fault localization, program comprehension, LLM agents, deterministic anchoring, repository navigation, SWE-bench}

\maketitle
\unmarkedfootnote{Accepted at ISSTA 2026.}

\section{Introduction}
\label{sec:introduction}

Large language models have transformed code assistance from single-completion
tools into autonomous agents that can navigate, analyze, and modify entire
repositories~\cite{chen2021,achiam2023gpt4}. Commercial agent systems such as Copilot Workspace, Claude Code, and
Gemini Code Assist~\cite{github2024copilotworkspace,anthropic2025claudecode,google2024geminicodeassist},
along with open research agents like SWE-agent and
SWE-Gym~\cite{yang2024sweagent,pan2024swegym}, have converged on a shared
architecture: \emph{grep-first retrieval} paired with strong LLM reasoning.
Given a natural language description of a bug or feature request, the agent
iteratively issues keyword queries (\url{grep}/\url{rg}), inspects
matching snippets, refines its hypotheses, and searches again. This design is
attractive because it is fast, language-agnostic, robust to broken code, and
requires no heavyweight indexing infrastructure. Empirically, these grep-first
agents are highly competitive and often outperform more complex graph-based
approaches in end-to-end repair benchmarks~\cite{yang2024sweagent,jimenez2024swebench,xia2022lesstraining,li2025giantrepair}.

Yet despite this practical success, grep-first agents suffer from a fundamental
\emph{representation mismatch}. Grep returns lines ranked by simple lexical
criteria and does not expose the \emph{structural relationships} that organize
real software: who calls whom, how configuration values propagate, which
classes inherit from which base, or how data flows between modules. Whether a
line is relevant to a bug depends not only on its keyword overlap with a query
but also on where it sits in this structural topology. This mismatch has two critical
consequences. First, agents must rediscover structural links through ad-hoc
multi-hop queries, resulting in fragmented, locally myopic context. Second,
LLM controllers are inherently stochastic: small perturbations in early searches
cascade into qualitatively different navigation paths. When coupled with
structure-blind retrieval, this produces \emph{brittle trajectories}: two runs
on the same task may visit different files, take different numbers of steps,
and produce different patches, even when the final outcome is identical.
Recent work on agent trajectories similarly highlights the sensitivity of
tool-using agents to exploration decisions and early branching~\cite{song2024eto}.

From a software engineering perspective, this \emph{behavioral unpredictability} is a critical quality concern: practitioners need agent behavior to be inspectable and reproducible, not merely correct in aggregate~\cite{angermeir2025reproducibility,baltes2025guidelines}. Our goal is therefore to understand the \emph{marginal value of static structure} when added to already-strong grep-first agents. We start from a baseline that already outperforms graph-based alternatives by a wide margin (Section~\ref{sec:motivation}), ask what additional benefit lightweight structural annotations provide, and treat trajectory quality and stability as first-class outcomes alongside localization accuracy.

This goal leads to a key insight: improving grep-first agents does not require abandoning text retrieval or building a new graph-guided controller. Instead, we can \emph{inject stable program structure directly into the agent's text view}. We call these structure facts \emph{deterministic anchors}: repository-level relationships (e.g., who calls whom, inheritance links, configuration usage) that are fixed for a given code snapshot and therefore should not vary across runs. Exposing them inline reduces the need for the agent to rediscover links via ad-hoc search and can make exploration more disciplined under stochastic LLM control.

\begin{figure*}[h]
    \centering
    \includegraphics[width=\linewidth]{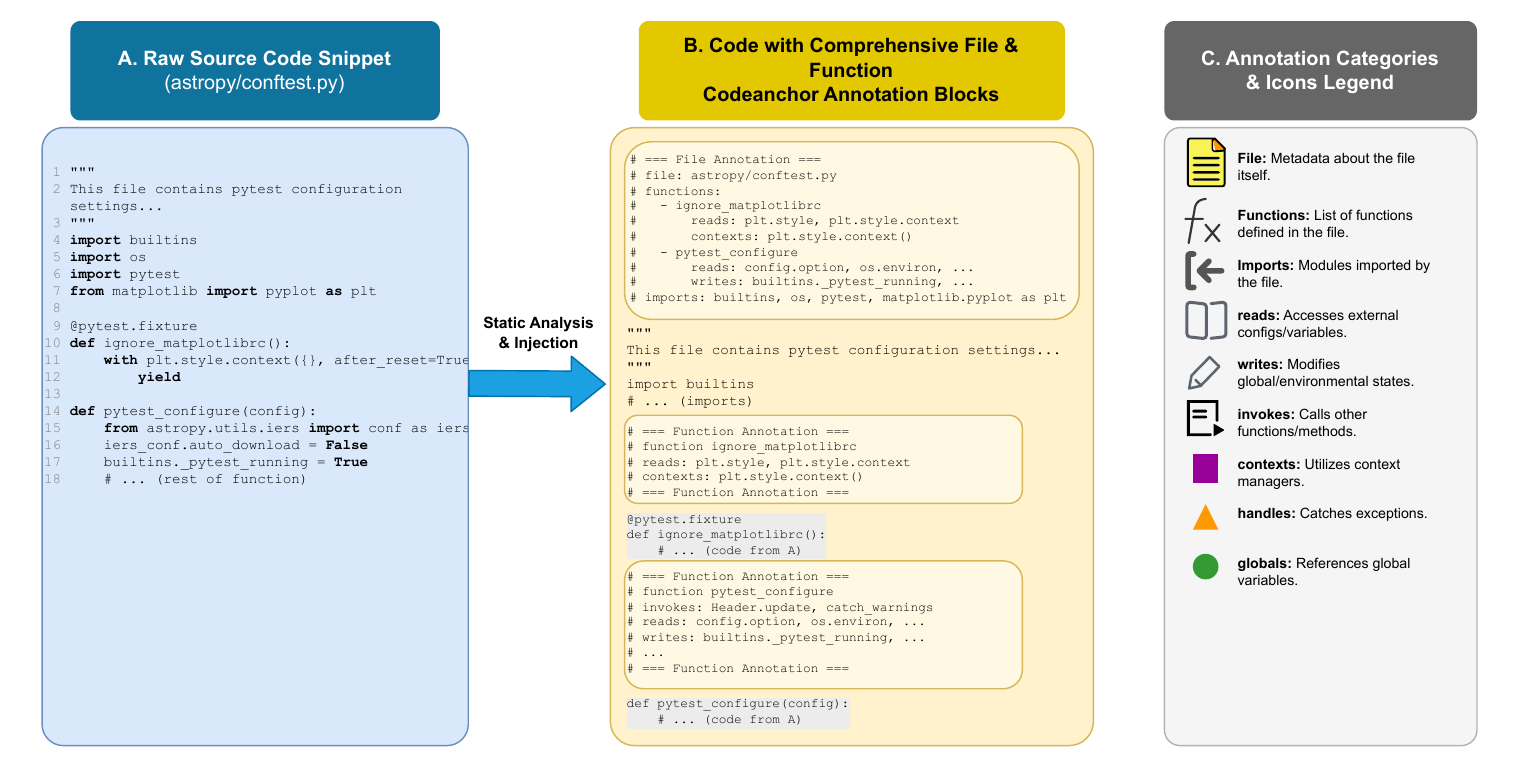}
    \caption{The framework of \ourapproach{}: Offline static analysis (PyCG call graphs + AST extractors) generates structured tags encoding call/inheritance/data-flow relationships, which are injected into source files as plain-text comments. At runtime, the grep-first agent retrieves code with embedded tags, enabling structure-guided navigation without changing the agent loop.}
    \label{fig:CodeAnchor}
\end{figure*}

To explore this idea, we present \ourapproach{}, a framework that augments
grep-based code agents with \emph{static-analysis-based structured comments}.
\ourapproach{} performs lightweight static analysis offline to extract
structural relationships such as calls, inheritance, data flow, and
configuration usage. It then injects these facts back into the codebase as
compact ``CodeAnchor tags,'' which are plain-text comments colocated with functions,
classes, and configuration entries. 
Figure~\ref{fig:CodeAnchor} shows an example of injecting tags into the code.
At runtime, the agent continues to issue
ordinary text searches over the repository, but search results now contain both
raw code and nearby tags that surface repository-internal cross-file structure
in place. Because tags are in-band text, they can influence both where the
agent lands and how it navigates after opening a file. This design enables what we call \emph{retrieval short-circuiting}: when the agent inspects an entity, the surrounding tags already reveal nearby structural links (e.g., callers/callees, imports, or configuration usage). The agent can follow these links directly instead of rediscovering them via additional searches, constraining exploration and making trajectories more consistent across stochastic runs.

We instantiate \ourapproach{} on top of Codex, a programmable code agent,
and evaluate it on SWE-bench Lite and SWE-bench
Verified~\cite{jimenez2024swebench}. We ask:
\begin{itemize}[leftmargin=*]
  \item \textbf{RQ1 (Does Topology Help?):} Does injecting basic topological structure (call/inheritance links) improve localization and interaction efficiency over pure keyword search?
  \item \textbf{RQ2 (Granularity and Directionality):} How do varying levels of semantic detail and link directionality affect agent behavior across different repository scales?
  \item \textbf{RQ3 (Behavioral Change \& Stability):} How do structural cues reshape trajectories, and to what extent do deterministic anchors improve run-to-run stability?
\end{itemize}
Across these questions, our primary lens is
\emph{behavioral}: we treat trajectory quality and stability as first-class
outcomes and interpret function-level localization gains as secondary evidence that
injected structure is genuinely used rather than ignored. We release the
implementation, tagging pipeline, and evaluation traces as open source at
\url{https://github.com/mathieu0905/Code-Anchor}.

In summary, our study identifies what we call the \textbf{deterministic anchoring effect}: static structure helps less by making agents ``smarter'' and more by making their navigation \emph{disciplined and reproducible}. Three observations support this finding: \bcircle{1}~\textbf{Anchoring works}: lightweight call/inheritance topology improves function-level localization (+2.2pp Func@5) and shortens trajectories ($-$1.6 interaction rounds) (Section~\ref{sec:rq1}); \bcircle{2}~\textbf{Anchoring is scale-sensitive}: optimal granularity and directionality depend on repository scale, with hub-heavy projects favoring inverse-only links (Section~\ref{sec:rq2}); \bcircle{3}~\textbf{Anchoring stabilizes}: tags raise link-following rate from 0.15--0.18 to 0.21--0.24 and roughly halve run-to-run variance (Section~\ref{sec:rq3}). \rev{These gains cost $\sim$9.9\% more input tokens on Lite, motivating practical guidelines:} default to lightweight topology on medium projects, prune forward edges in large repositories, and reserve dense tags for implicit-dependency cases.

\section{Background}
\label{sec:background}

This section introduces the grep-first agent architecture that serves as the foundation for our study.

\paragraph{Grep-first code agents.} Modern code agents~\cite{achiam2023gpt4,chen2021,yang2024sweagent} follow a tool-using pattern: given an issue description, the agent iteratively issues keyword queries (\url{grep}/\url{rg}), inspects matching snippets, refines its hypothesis, and searches again. Commercial assistants (Copilot Workspace, Claude Code, Gemini Code Assist)~\cite{github2024copilotworkspace,anthropic2025claudecode,google2024geminicodeassist} predominantly adopt this architecture for its simplicity, language-agnosticism, and robustness to incomplete code. The LLM is responsible for precision: filtering irrelevant matches, reasoning about code semantics, and deciding where to navigate next.

\paragraph{The connectivity gap.} While grep-based retrieval excels at high-recall keyword matching, it treats the codebase as an unstructured collection of text. Structural relationships that organize real software, such as who calls whom, which classes inherit from which base, and how configuration values propagate, remain implicit. Agents must rediscover these links through ad-hoc multi-hop queries, resulting in fragmented context and unpredictable navigation paths. This motivates our investigation: can lightweight structural annotations bridge this connectivity gap without abandoning the simplicity of grep-first retrieval?

\section{Motivation and Problem Analysis}
\label{sec:motivation}

For the agents we study, the core interaction pattern is a retrieve-then-read loop: the agent issues text queries against the repository, opens matching snippets, updates its hypothesis, and repeats. Tools such as \url{grep} or BM25~\cite{robertson2009bm25} treat the codebase as unstructured text and rank by lexical relevance. With modern reasoning models, this has become a hard-to-beat baseline: grep supplies broad candidates, the LLM supplies precision. What this view still lacks is \emph{connectivity}. Hybrid retrieval (e.g., BM25 + embedding reranking~\cite{lewis2020rag}) improves matching but still returns independent text spans without making call, inheritance, or configuration/data-flow links explicit. Yet whether a location is relevant to a bug depends on how it sits in the program's structure---which configuration values reach it, which functions it calls or overrides, and which downstream components it affects.

\begin{figure*}[h]
    \centering
    \includegraphics[width=\linewidth]{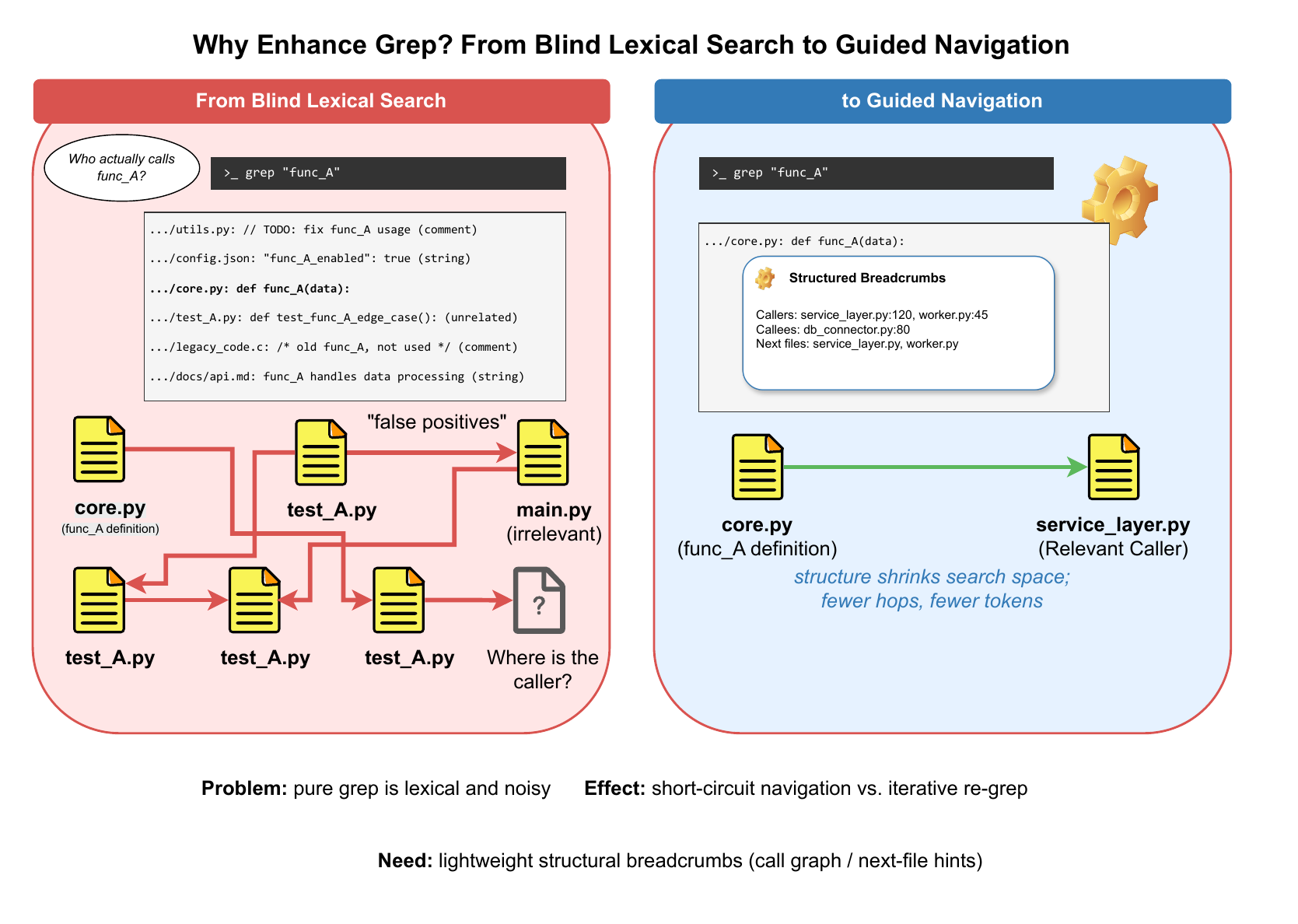}
    \caption{Motivating example: pure grep-style keyword search is lexical and noisy, while inline CodeAnchor tags surface structured breadcrumbs (e.g., callers/callees and next-file hints) to guide navigation.}
    \label{fig:motivation-codeanchor}
\end{figure*}

Consider a configuration-driven bug as a running example (shown in Figure~\ref{fig:motivation-codeanchor}). Changing a single
timeout in a YAML file may indirectly affect a database connection pool, a
circuit breaker, and a load balancer, each implemented in different modules. An
engineer investigating such an incident usually follows a structural chain:
identify the configuration, find all its uses, understand how the value is
transformed, and reason about the downstream components. A structure-blind
agent, in contrast, must rediscover each link via additional searches. The
result is a fragmented, locally myopic view: the agent may see strong textual
evidence at a call site yet never reach the internal helper where the actual
fix belongs.
This perspective closely mirrors classic change impact analysis workflows that
trace dependency chains across modules~\cite{bohner1996software,li2020survey,kretsou2021change,borg2017software,lawall2016interprocedural}.

This structural mismatch interacts with the stochastic nature of LLM
controllers. An agent run is not a single response but a \emph{trajectory} that
interleaves tool calls and model invocations. Small differences in early search
queries or intermediate summaries can send the agent down different branches of
the structural chain. Our preliminary observations (later formalized in RQ3)
indicate that, on the same SWE-bench Lite issue and with the same agent
configuration, repeated runs can visit different sets of files, take different
numbers of search steps, and produce different patches, even when the final
pass/fail outcome is the same. For practitioners, this means that behavior is
difficult to predict or reproduce, and failures are hard to debug
because different runs often fail for different structural reasons~\cite{song2024eto,angermeir2025reproducibility}. 
\rev{In real deployments, developers only run the agent once. However, the trajectory variance leads to the situation that a bug found in one run may be missed in the next. Engineers cannot tell whether the failure reflects a systematic weakness or just stochastic noise. If agents can run stable, the failures may be more interpretable: when two failed runs follow the same path, the missing step is plain and can be addressed; when every run fails differently, each failure is a fresh investigation.} Therefore, in the rest of this section, we treat fragility and run-to-run variance as symptoms of the underlying mismatch between text-oriented retrieval and structure-oriented code, and ask how much explicit structure is needed to control them.

Our goal is not only to make code agents smarter but also to make their behavior less random. Source code contains rich \emph{deterministic structure}---call graphs, inheritance hierarchies, data-flow and configuration usage, module-level dependencies---that does not change across runs for a fixed repository. Surfacing such facts inline lets them serve as \emph{deterministic anchors} for an otherwise stochastic process. This leads to \ourapproach{}: run lightweight static analysis offline, inject the resulting structural facts back into the codebase as compact structured comments (CodeAnchor tags), and let the agent continue using its familiar grep-based retrieval over both raw code and these annotations. When the agent opens a function, the surrounding tags already reveal its callers, call targets, and upstream configuration, short-circuiting the multi-hop searches that would otherwise be needed. We keep the underlying agent loop, tools, and model fixed, and study this design on SWE-bench Lite and Verified~\cite{jimenez2024swebench}.

A natural alternative is to expose structure via an explicit MCP \rev{(Model Context Protocol)} tool (e.g., LSP ``find references''~\cite{microsoft2016lsp}). However, in a pilot on 20 SWE-bench tasks with an optional call-graph tool, we observed low tool-use rates: the agent typically relied on plain \url{grep} instead. This motivates our ``passive injection'' stance: surface structural context inline, right next to the code the agent opens, so that following structure does not require an explicit tool-usage decision. We also choose to strengthen a grep-first loop rather than build a new graph-guided controller because our goal is to study the marginal impact of static structure on architectures that practitioners already deploy. To validate this focus, we benchmarked LocAgent~\cite{chen2025locagent} on SWE-bench Lite under matched conditions: the graph agent trails grep-first by 23.7pp (59.5\% vs.\ 83.2\% \rev{Func@5: top-5 predictions cover all ground-truth functions; see Section~\ref{sec:evaluation} for full definitions of File@$k$ and Func@$k$}, Table~\ref{tab:prelim-locagent}).

\begin{table*}[h]
\centering
\caption{Preliminary SWE-bench Lite head-to-head (matched model/limits, $N{=}274$).}
\label{tab:prelim-locagent}
\begin{tabular}{lccccc}
\toprule
Agent & File@1 & File@3 & File@5 & Func@5 & Func@10 \\
\midrule
LocAgent (graph-based) & 0.726 & 0.850 & 0.861 & 0.595 & 0.664 \\
Codex (grep-based)     & 0.912 & 0.967 & 0.967 & 0.832 & 0.836 \\
\bottomrule
\end{tabular}
\end{table*}

Analysis reveals clear grep-based failure modes: \emph{function-level failures} where the agent stops at call sites without reaching private implementations. Textual evidence finds the right file, but disambiguating similarly named functions requires explicit knowledge of who calls whom. These observations motivate our design: keep grep-first tools fixed, and inject just enough static structure to repair multi-hop and disambiguation failures.

Based on this analysis, the next section presents \ourapproach{}, our concrete instantiation of deterministic anchors for grep-based code agents.

\section{Approach}
\label{sec:approach}

We now describe \ourapproach{}, our instantiation of static
structure-as-anchors for grep-based code agents. We first provide an overview of the system,
then detail how we build CodeAnchor tags via lightweight static analysis, and finally
explain how these tags integrate with an existing Codex-style agent. To make the
design concrete, we also present a pseudocode description of the offline tag
generation pipeline.

\subsection{System Overview}

\ourapproach{} has two components: an \emph{offline static analysis pipeline} that extracts structural relationships from the codebase and encodes them as structured comments (CodeAnchor tags), and an \emph{agent integration layer} that lets a grep-driven agent consume these tags without changing its high-level control loop. The offline pipeline runs once per repository snapshot (or incrementally after code changes) and is \emph{task-agnostic}: it depends only on repository contents and does not use the natural-language issue description or any ground-truth localization/patch metadata. It parses source files, identifies key entities (functions, classes, configuration entries, files), and computes relationships such as callers, callees, inheritance, imports, containment, data dependencies, and configuration usage. These relationships are attached to entities as synthetic comments in a compact, machine-parsable format.

At runtime, the agent continues to use plain text search (\url{rg}) over the working tree. Because CodeAnchor tags are inserted directly into source files, search results now include both code and nearby tags, which the agent reads as ordinary text in its context window. No new tools or APIs are required beyond grep and file I/O. In particular, we do not add a separate embedding index or an explicit reranking stage: retrieval remains the same grep-first primitive, and only the repository view changes.

\subsection{CodeAnchor Tags: Static-Analysis-Based Structured Comments}

CodeAnchor tags are structured comment blocks colocated with code entities. They follow a simple schema:
\begin{quote}
\begin{verbatim}
# === Function Annotation ===
# function _get_locale_dirs
# used by: fetch, lang_stats, update_catalogs
# === Function Annotation ===
def _get_locale_dirs(resources, include_core=True):
\end{verbatim}
\end{quote}
The exact syntax varies across languages (e.g., \url{//} for Java/C++, \url{\#} for Python), but the semantics are uniform: each tag block has a stable identifier for the entity it annotates (\url{function}, \url{class}, or \url{file}), fields encode structural relationships (\url{used by}, \url{invokes}, \url{inherits}, \url{imports}), and values are normalized references to other entities. Because tags are structured comments with uniform delimiters (\url{=== Annotation ===}), they are trivially identifiable and removable via regular-expression passes, ensuring that repositories can cleanly revert to untagged code if needed.

We distinguish two tiers of relationships that correspond to the two configurations in our experiments. In the \url{Anchor-Topo} configuration, we surface only four LocAgent-style structural relations~\cite{chen2025locagent}: \emph{contains}, \emph{imports}, \emph{invokes}, and \emph{inherits}. These appear in tags as fields such as \url{PARENT}/\url{CHILDREN} (for containment), \url{IMPORTS}/\url{IMPORTED\_BY} (for module imports), \url{CALLS}/\url{CALLED\_BY} (for function calls), and \url{BASE}/\url{DERIVED} (for class inheritance). This lightweight topology gives the agent an explicit call graph and class hierarchy, plus the file/module skeleton, without attempting to encode richer semantics.

In the \url{Anchor-Dense} configuration we add further relations on top of this core: configuration-driven edges such as \url{CONFIG\_USAGE}, \url{ENV\_VAR}, and \url{CONST\_REF} that link config entries and constants to the code that consumes them; simple data-flow edges such as \url{DATA\_DEP} and \url{IO\_DEP} that sketch where values and I/O effects travel~\cite{weiser1984,horwitz1990,tip1995survey,korel1988}; and a small number of domain-specific links (e.g., test-to-code mappings and plugin registrations) that help in repositories with heavy plugin or test harness use. These extra fields are exactly what distinguishes \url{Anchor-Dense} from \url{Anchor-Topo} in RQ2: both share the same four structural backbones, but only the full variant carries configuration and data-flow hints.

Finally, we vary directionality to study how much ``forward'' navigation signal is useful. The \url{Anchor-Inv} configuration keeps only inverse dependencies such as \url{CALLED\_BY} (dropping \url{CALLS}). \rev{This is not suggested as a universal improvement over the policy; it isolates one type of tension that arises with hub-heavy projects when links to functions with high degrees repeat the name of the function throughout many call sites, allowing that hub to eclipse others in keyword search. Inverse-only tags keep backward navigability at this expense.} In principle, intermediate policies are possible (e.g., degree caps or hub filtering on forward links), but we focus on these clean variants to isolate behavioral effects under a fixed grep-first loop.

\paragraph{Scalability and Language Support.}
The analysis is language-pluggable: the tag schema and agent integration are language-agnostic, while relationship extractors can be implemented with language-specific tooling. For established languages such as Java, JavaScript, C++, and Go, mature static-analysis frameworks (e.g., WALA, Soot, CodeQL, Semgrep, or LLVM-based tools) can provide call graphs and dataflow information~\cite{wala2024,soot1999,lattner2004llvm,codeql2019,semgrep2020}. For
newer or less-supported languages, tree-sitter~\cite{tree-sitter} enables
rapid development of lightweight AST-based extractors that capture
containment, imports, and simple call patterns without requiring a full
compiler frontend. For repositories where deep interprocedural analysis
is infeasible, \ourapproach{} still provides value via the four structural
relations alone, which is precisely the setting captured by our
\url{Anchor-Topo} runs, in line with classic guidance on scalable static
analysis~\cite{cousot1977abstract}.

\subsection{Offline Tag Generation}

The offline pipeline constructs CodeAnchor tags for a given repository snapshot by: (1) parsing all source files into an intermediate representation, (2) identifying entities (functions, classes, config entries, files), (3) for each entity, inferring call relations, data flow, and domain links using language-specific frontends, (4) normalizing relationships to stable identifiers, (5) rendering tags as language-specific comments, and (6) inserting tags near definition sites. This workflow mirrors classic dependency extraction used in impact analysis and regression testing~\cite{orso2003leveraging,binkley2007application}.

\rev{Two terms above need clarifying. A \emph{stable identifier} is how we name a code entity inside a tag; for a function, we use its file path with its dotted name (e.g., \url{django/db/backends/base/schema.py:BaseDatabaseSchemaEditor._delete_composed_index}), so tags survive whitespace, reformatting, or moves within a file, while renames and cross-file moves regenerate the tag. \emph{Domain links} are the extra relationships added by \url{Anchor-Dense} on top of call/inheritance: \url{CONFIG_USAGE} for code reading Django/Flask settings, \url{CONST_REF} for code using module-level constants, \url{IO_DEP} for functions touching files, network, or databases, and \url{TEST_REF} for test files matched to production modules; all four come from AST pattern matches rather than interprocedural analysis, keeping the pipeline fast and free of spurious edges.}
\label{sec:cost}
\rev{Tag generation is a one-time cost paid per repository snapshot and reused across all queries. On four Python repositories, static graph construction (AST walk plus call/inheritance/import extraction) takes \textbf{6.8\,s} (\url{pytest-7432}, 73k LOC), \textbf{21.1\,s} (\url{sklearn-15512}, 247k), \textbf{58.4\,s} (\url{astropy-12907}, 341k), and \textbf{133.4\,s} (\url{django-13658}, 367k LOC); cost scales near-linearly with LOC and graph size. Incremental regeneration after code changes touches only the affected files.} 


\paragraph{Python Instantiation.}
Our prototype targets Python using PyCG~\cite{salis2021pycgpracticalgraphgeneration} for call graphs and AST passes for containment/imports/inheritance (\url{Anchor-Topo} config). For \url{Anchor-Dense}, we add lightweight intraprocedural extractors for config usage and data/I/O hints. All analysis prioritizes precision over completeness: PyCG and our syntactic extractors emit only verifiable relationships, avoiding spurious edges from dynamic dispatch, reflection, or monkey patching, which are well-known sources of unsoundness in static analysis~\cite{bodden2011taming}. We accept incomplete coverage for lightweight, CI-deployable pipelines; results (Section~\ref{sec:evaluation}) show that the captured relationships suffice for most localization tasks.

\subsection{Agent Integration}

We integrate \ourapproach{} with a Codex-style agent exposing text search, file open, patch application, and test execution. Crucially, we do \emph{not} modify the agent's outer control loop; tags are treated as part of the codebase. The agent issues keyword searches (hitting both code and tags), opens files and reads nearby tags as hints, and follows explicit links or issues new searches. This realizes \emph{retrieval short-circuiting}: tags provide deterministic local structure, reducing search calls and constraining plausible next steps.

We compare four retrieval views: Baseline (raw grep), \url{Anchor-Topo} (bidirectional call/inheritance), \url{Anchor-Dense} (adds data-flow/config edges), and \url{Anchor-Inv} (inverse-only). The localization prompt is purely task-oriented and makes no reference to tags; the model treats them as ordinary comments.

\section{Evaluation}
\label{sec:evaluation}

We evaluate \ourapproach{} by integrating it into a Codex-style programmable
code agent and running it on SWE-bench Lite and SWE-bench Verified~\cite{jimenez2024swebench}, which are benchmarks of
real GitHub issues paired with tests. This agent architecture matches the
grep-first design used by many modern industrial assistants and is already
highly competitive: on SWE-bench Lite, for example, our raw baseline achieves
Func@5 $\approx$83.2\%. Our evaluation is structured around the three research
questions introduced in Section~\ref{sec:introduction}:
\begin{itemize}
    \item \textbf{RQ1: Does Topology Help?} - Does injecting basic topological structure (call and inheritance links) improve localization accuracy and interaction efficiency over pure keyword search?
    \item \textbf{RQ2: Granularity and Directionality} - How do varying levels of semantic detail (basic topology vs.\ dense annotations) and link directionality (bidirectional vs.\ inverse-only) affect agent behavior across different repository scales?
    \item \textbf{RQ3: Behavioral Change \& Stability} - How does structural injection alter the agent's navigation trajectory, and to what extent can deterministic anchors mitigate the stochastic nature of LLM navigation across repeated runs?
\end{itemize}

\subsection{Experimental Setup}

\paragraph{Model and agent.}
We use the \url{Codex} coding agent with OpenAI's GPT-5.1-codex model (thinking effort set to high). The agent exposes many bash commands such as \textsc{Search} (\url{rg}), \textsc{Open}, \textsc{ApplyPatch}, and \textsc{RunTests}; we focus on localization and disable patching/tests to isolate navigation effects.

\paragraph{Prompt.}
All configurations share the same localization prompt template, which is purely task-oriented and makes no reference to CodeAnchor tags or any tag-following policy. Thus, the only experimental variable is whether tags are present in the retrieved code context. The prompt instructs the agent to localize files, classes, and functions for a given GitHub issue and return a JSON response with ranked candidates sorted by confidence.

\paragraph{Benchmark configuration.}
We use SWE-bench Lite and Verified~\cite{jimenez2024swebench}, and for SWE-bench Lite, we use the same 274 instances following the setting of LocAgent. For the stability analysis in RQ3, we construct a 50-instance subset per dataset with $k{=}10$ repeats, prioritizing boundary cases where Baseline and \url{Anchor-Topo} disagree, then filling with uniform random samples.

\paragraph{Configurations.}
We compare four retrieval views of the same agent: a Baseline over raw code, an \url{Anchor-Topo} version with forward and backward call/inheritance tags, an \url{Anchor-Dense} variant that also injects data-flow and configuration edges, and an inverse-only \url{Anchor-Inv} variant that retains only ``who-calls-me'' links (no forward call edges). RQ1 focuses on Baseline vs.\ \url{Anchor-Topo} to establish whether structure helps at all; RQ2 compares the three structural variants (\url{Anchor-Topo}, \url{Anchor-Dense}, \url{Anchor-Inv}) to explore granularity and directionality effects; RQ3 examines behavioral changes primarily through Baseline vs.\ \url{Anchor-Topo} to isolate stability effects under a fixed topology choice.

\paragraph{Metrics.}
We evaluate localization performance using File@$k$ and Func@$k$. A task is considered successful if and only if the top-$k$ retrieved entities (files or functions) encompass the entire set of ground-truth targets for that task; the metric reports the percentage of such successful tasks across the dataset.
\rev{On \emph{Lite} (274 instances), every instance has one ground-truth file. 84.0\% of them have a single ground-truth function (230/274); 11.3\% have two, and the remaining 4.7\% have three or more. \emph{Verified} (500 instances) is harder: 85.8\% have one ground-truth file and 14.2\% have two or more; 64.2\% have a single ground-truth function and 35.8\% have two or more. Func@$k$ at larger $k$ therefore tests whether the top-$k$ predictions cover all ground-truth functions, which is most demanding on multi-function tasks.}

We also report Rounds, defined as the average number of tool calls per task, to measure interaction efficiency. \rev{A \emph{navigation transition} is any tool call that opens, searches, or inspects a code entity; Hops counts all such transitions summed across tasks.} Link Following Rate (LFR) measures trajectory guidance: for each navigation step $t \to t+1$, we classify the transition as \emph{structural} if the opened entity at $t+1$ was mentioned in a CodeAnchor tag visible at $t$, \emph{lexical} if it appeared in non-tag text at $t$, or \emph{exploratory} otherwise. LFR is the fraction of steps that are either structural or lexical (i.e., non-exploratory). Struct and Lex denote the structural-only and lexical-only fractions, respectively. We quantify uncertainty via nonparametric bootstrap, and report stability as mean $\pm$ std across 10 runs.

\subsection{RQ1: Does Topology Help?}
\label{sec:rq1}

\textbf{RQ1}: \emph{Does injecting basic topological structure (call and inheritance links) improve localization accuracy and interaction efficiency over pure keyword search?}

Before exploring richer annotations or alternative configurations, we first establish whether structural topology provides measurable benefit at all. We compare Baseline (raw grep over untagged code) against \url{Anchor-Topo} (bidirectional call/inheritance tags) on SWE-bench Lite (274 tasks) and Verified (500 tasks). Table~\ref{tab:rq1} summarizes results.

\begin{table*}[h]
\centering
\caption{RQ1: Baseline vs.\ \url{Anchor-Topo} on SWE-bench Lite and Verified.}
\label{tab:rq1}
\scalebox{0.85}{
{\small
\setlength{\tabcolsep}{4pt}
\renewcommand{\arraystretch}{1.1}
\begin{tabular}{@{}lccccc|ccccc@{}}
\toprule
& \multicolumn{5}{c|}{SWE-bench Lite} & \multicolumn{5}{c}{SWE-bench Verified} \\
Config & File@1 & File@3 & Func@5 & Func@10 & Rounds & File@1 & File@3 & Func@5 & Func@10 & Rounds \\
\midrule
Baseline &
\textbf{0.9124} & 0.9672 & 0.8321 & 0.8358 & 35.3 &
0.8520 & 0.8680 & 0.6187 & 0.6227 & 42.4 \\
\url{Anchor-Topo} &
0.8978 & \textbf{0.9745} & \textbf{0.8540} & \textbf{0.8577} & \textbf{33.7} &
0.8480 & 0.8620 & \textbf{0.6308} & \textbf{0.6369} & \textbf{40.9} \\
\midrule
$\Delta$ & $-$1.5pp & +0.7pp & \textbf{+2.2pp} & \textbf{+2.2pp} & \textbf{$-$1.6} & $-$0.4pp & $-$0.6pp & \textbf{+1.2pp} & \textbf{+1.4pp} & \textbf{$-$1.5} \\
\bottomrule
\end{tabular}
}
}
\end{table*}

\paragraph{Localization Accuracy.}
On Lite, \url{Anchor-Topo} improves function-level recall by +2.2pp (Func@5: 0.8540 vs.\ 0.8321; Func@10: 0.8577 vs.\ 0.8358), with the Func@5 improvement reaching statistical significance (McNemar $p{=}0.041$). On Verified, improvements are consistent though smaller (+1.2pp Func@5, +1.4pp Func@10; McNemar $p{=}0.023$ for Func@10). The baseline already achieves 83.2\% Func@5 on Lite, leaving limited headroom; the observed gains therefore reflect improvements in the regime of an already strong agent.

\paragraph{Interaction Efficiency.}
Tags also reduce trajectory length: $-$1.6 rounds on Lite (33.7 vs.\ 35.3) and $-$1.5 rounds on Verified (40.9 vs.\ 42.4). Shorter trajectories suggest that explicit structural links help agents reach targets more directly, reducing exploratory tool calls.

\paragraph{Structural Detour.}
We observe a notable pattern: \url{Anchor-Topo} slightly lowers File@1 on Lite ($-$1.5pp) even while improving Func@10. This reflects what we term a \emph{structural detour}: agents with tags may first visit hub or helper files surfaced by structural annotations, then navigate to the ground-truth function within the top few opens. Trace analysis shows that a substantial fraction (34\% of Verified tasks exhibiting File@1 regression) still reach the target function within 3 opens despite initially visiting a structural neighbor. In essence, agents trade an immediate ``lucky'' lexical hit for a grounded structural path, a trade-off that pays off in final function identification.

\paragraph{Downstream Impact: Localization $\to$ Repair.}
\rev{To test whether localization gains propagate to repair, we isolate the 80 SWE-bench Verified instances where Baseline and \url{Anchor-Topo} disagree on the top-5 function set, then run a controlled repair experiment on this differential subset: the repair agent is held fixed (GPT-5.1-codex), and only the top-5 localized functions vary. \url{Anchor-Topo} resolves 48/80 against Baseline's 38/80 (60.0\% vs.\ 47.5\%, +12.5\,pp), with a strict \emph{superset property}: every Baseline-repaired instance is also repaired by \url{Anchor-Topo}, plus 10 additional cases (zero Baseline-exclusive repairs). This is a conditional effect $\mathbb{E}[\Delta\text{Fix} \mid \Delta\text{Loc}>0]$; projected to all 500 Verified instances, the 10 extra repairs correspond to $\sim$+2\,pp overall, consistent with the +1.2\,pp Func@5 gain.}

\answer{1}{
\begin{itemize}[leftmargin=*, nosep]
  \item Structural topology reliably improves function-level recall (\textbf{+2.2pp Func@5} on Lite, \textbf{+1.2pp} on Verified) over a strong grep baseline.
  \item Tags shorten trajectories (\textbf{$-$1.6 rounds} on Lite, \textbf{$-$1.5 rounds} on Verified), indicating more direct navigation.
  \item Structural detours may lower early file hits but ultimately improve function identification.
  \item Localization gains propagate to repair on the differential subset: \textbf{+12.5pp} repair success with a strict superset property; projected to full Verified, $\sim$+2\,pp overall.
\end{itemize}
}

\subsection{RQ2: Granularity and Directionality}
\label{sec:rq2}

\textbf{RQ2}: \emph{How do varying levels of semantic detail (basic topology vs.\ dense annotations) and link directionality (bidirectional vs.\ inverse-only) affect agent behavior across different repository scales?}

Having established that structural injection benefits localization (RQ1), we now investigate how different structural configurations affect agent behavior. We explore two dimensions: (1)~\textbf{granularity}, whether denser semantics (data-flow, configuration edges) beyond basic topology provide additional benefit or introduce overhead; and (2)~\textbf{directionality}, how bidirectional vs.\ inverse-only links interact with repository characteristics.

We compare three structural configurations: \url{Anchor-Topo} (bidirectional call/inheritance), \url{Anchor-Dense} (adds data-flow and config edges), and \url{Anchor-Inv} (inverse-only, i.e., ``who-calls-me'' links without forward edges). Table~\ref{tab:rq2} summarizes results.

\rev{In RQ2, we combine \url{call} and \url{inheritance} into one ``basic topology'' type. For instance, call edges have roughly 8--20$\times$ than inheritance edges in our Python repositories; an inheritance-only variant would be dominated by a few class-hierarchy cases while failing to have enough power under our $\alpha{=}0.05$ design to detect differences. Granularity and Directionality are then the dimensions where effect sizes stay large enough to discriminate, which is what RQ2 studies.}

\begin{table*}[h]
\centering
\caption{RQ2: Effect of granularity and directionality on localization and efficiency. \rev{Tokens are mean per-instance input tokens; Baseline shown for reference. }}
\label{tab:rq2}
\scalebox{0.7}{
\begin{tabular}{lccccc|ccccc}
\toprule
& \multicolumn{5}{c|}{SWE-bench Lite} & \multicolumn{5}{c}{SWE-bench Verified} \\
Config & Func@5 & Func@10 & Rounds & \rev{Input tok.} & \rev{$\Delta$ vs.\ Base} & Func@5 & Func@10 & Rounds & \rev{Input tok.} & \rev{$\Delta$ vs.\ Base} \\
\midrule
Baseline & 0.8321 & 0.8358 & 35.3 & 406k & --- & 0.6187 & 0.6227 & 42.4 & 567k & --- \\
\url{Anchor-Topo} & \textbf{0.8540} & \textbf{0.8577} & \textbf{33.7} & 446k & +9.9\% & 0.6308 & 0.6369 & \textbf{40.9} & 618k & +9.0\% \\
\url{Anchor-Dense} & \textbf{0.8540} & \textbf{0.8577} & 38.6 & 530k & +30.5\% & 0.6288 & \textbf{0.6389} & 41.6 & 647k & +14.1\% \\
\url{Anchor-Inv} & 0.8242 & 0.8278 & 55.0 & 566k & +39.4\% & \textbf{0.6329} & \textbf{0.6389} & 41.7 & 620k & +9.3\% \\
\bottomrule
\end{tabular}
}
\end{table*}

\paragraph{Effect of Granularity: Topo vs.\ Dense.}
Moving from \url{Anchor-Topo} to \url{Anchor-Dense} produces different effects across datasets. On Lite, function accuracy remains unchanged (Func@5/10 both 0.8540) while rounds increase by +4.9 and input-token usage grows by +18.8\% (530k vs.\ 446k). On Verified, \url{Anchor-Dense} marginally improves Func@10 (0.6389 vs.\ 0.6369) but slightly lowers Func@5 (0.6288 vs.\ 0.6308). These results indicate a \textbf{saturation effect}: most benefit comes from basic topology, and adding dense data/config edges yields diminishing returns while increasing context overhead.

Qualitatively, \url{Anchor-Dense} rescues a small set of long-tail implicit-dependency tasks: 3/274 Lite instances and 15/500 Verified instances involving multi-hop value propagation (e.g., Django encoding paths, Matplotlib collection state, SymPy shape propagation). However, full traces also reveal a trade-off: dense tags can shift attention toward high-degree helper functions, changing salience ordering and causing agents to explore utility modules before reaching the actual bug site.

\paragraph{Effect of Directionality: Bidirectional vs.\ Inverse-Only.}
Directionality effects are \textbf{scale-dependent}. On Lite (medium-scale repositories, mean 35k LOC), removing forward edges (\url{Anchor-Inv}) degrades performance: Func@5 drops to 0.8242 ($-$3pp vs.\ \url{Anchor-Topo}) and rounds inflate to 55.0 (+21.3 vs.\ \url{Anchor-Topo}). Without forward links, agents lose explicit ``where-to-go-next'' guidance and resort to more expensive keyword exploration.

On Verified (larger repositories, mean 120k LOC, 23\% hub nodes), the pattern differs. \url{Anchor-Inv} matches or exceeds other variants on function metrics (Func@5: 0.6329, Func@10: 0.6389) with an input-token footprint comparable to \url{Anchor-Topo} (620k vs.\ 618k). In hub-heavy projects, forward links from high-degree nodes generate repetitive tag text that can overexpose structurally central helpers in search results. Inverse-only tags surface ``who depends on me'' relationships without this amplification effect.

\vspace{1em}
\answer{2}{
\begin{itemize}[leftmargin=*, nosep]
  \item \textbf{Granularity saturation}: Dense data-/config-flow tags show diminishing returns (\textbf{+4.9 rounds} on Lite) beyond basic topology, though they rescue \textbf{3/274 Lite, 15/500 Verified} implicit-dependency cases.
  \item \textbf{Scale-dependent directionality}: Medium-scale repos benefit from bidirectional links; large hub-heavy repos show better results with inverse-only view (\textbf{+0.2pp Func@5} on Verified, at input-token parity with \url{Anchor-Topo}).
  \item \textbf{Practical implication}: Structural configuration should adapt to repository characteristics: lightweight topology for most cases, inverse-only for large projects, dense tags for implicit-flow debugging.
\end{itemize}
}

\subsection{RQ3: Behavioral Change \& Stability}
\label{sec:rq3}

\textbf{RQ3}: \emph{How does structural injection alter the agent's navigation trajectory, and to what extent can deterministic anchors mitigate the stochastic nature of LLM navigation across repeated runs?}

Beyond aggregate accuracy, we look at how agents actually use structural information during navigation, and whether static structure reduces the variance that comes with LLM-based navigation. The latter matters as much as average accuracy for production deployment.

\paragraph{Trajectory Behavior.}
We measure Link Following Rate (LFR): the fraction of navigation steps that follow an entity mentioned in a previous tag. Results from instrumented traces appear in Table~\ref{tab:trajectory}.

On Lite, structural tags raise LFR substantially over Baseline. \url{Anchor-Topo} reaches the highest overall and structural LFR (0.236 and 0.163), with \url{Anchor-Dense} close behind (0.212 / 0.158). \url{Anchor-Inv} drops to 0.186 overall and 0.143 structural, only slightly above Baseline's 0.178---without forward edges, the agent falls back to lexical hopping. This matches RQ2's finding that medium-scale repositories benefit from bidirectional links.

On Verified, all tagged configurations raise LFR (0.147 $\to$ 0.174--0.212). \url{Anchor-Topo} and \url{Anchor-Dense} lead (0.209 / 0.212), and \url{Anchor-Inv} sits between them (0.174, structural 0.096). High LFR is not always good on hub-heavy repositories: \rev{dense tags pull the agent into utility modules whose names show up everywhere but which have nothing to do with the bug---a \emph{structural distraction}, where the structural cue and the actual target point in different directions. \url{Anchor-Inv} avoids this because it has fewer links to chase in the first place.}

\rev{To check whether tags actually \emph{guide} the agent or just get followed mechanically, we compute the \emph{effective tag rate}: out of all hops the agent takes by following a tag, the fraction that lands on a ground-truth entity. This rate is 27.1\% on Lite and 26.6\% on Verified, and at the instance level, 55.2\% (Lite) and 58.7\% (Verified) of tasks contain at least one such hop. Tags are not merely followed; they reach the right target often enough to explain the accuracy gains in RQ1 and RQ2.}

Figure~\ref{fig:rq3-target-prob} backs this up visually: tagged configurations sustain higher target-hit probability across early steps on both datasets.

\begin{table*}[h]
\centering
\caption{Trajectory guidance metrics (RQ3). Hops = total navigation transitions across all tasks. Struct/Lex are structural-/lexical-link following rates; LFR is overall link-following rate.}
\label{tab:trajectory}
\small
\begin{tabular}{lcccc|lcccc}
\toprule
\multicolumn{5}{c|}{SWE-bench Lite} & \multicolumn{5}{c}{SWE-bench Verified} \\
Config & Hops & Struct & Lex & LFR & Config & Hops & Struct & Lex & LFR \\
\midrule
Baseline & 1132 & 0.000 & 0.178 & 0.178 & Baseline & 2426 & 0.000 & 0.147 & 0.147 \\
\url{Anchor-Topo} & 1059 & \textbf{0.163} & 0.162 & \textbf{0.236} & \url{Anchor-Topo} & 2521 & \textbf{0.153} & 0.144 & 0.209 \\
\url{Anchor-Dense} & 1200 & 0.158 & 0.154 & 0.212 & \url{Anchor-Dense} & 2483 & 0.149 & 0.151 & \textbf{0.212} \\
\url{Anchor-Inv} & 586 & 0.143 & 0.154 & 0.186 & \url{Anchor-Inv} & 2411 & 0.096 & 0.140 & 0.174 \\
\bottomrule
\end{tabular}
\end{table*}

\begin{figure}[h]
\centering
\includegraphics[width=\linewidth]{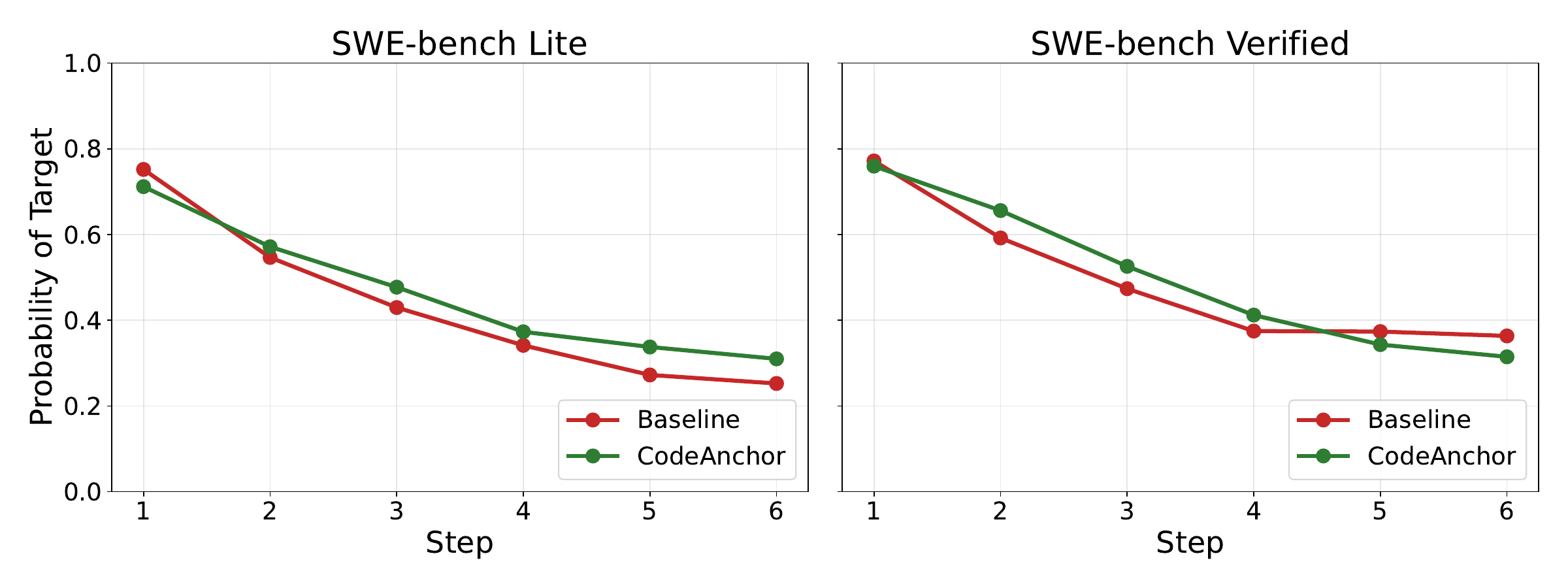}
\caption{Target-file hit probability over steps (RQ3). CodeAnchor tags sustain higher early hit rates on both datasets.}
\figlabel{rq3-target-prob}
\end{figure}

\paragraph{Run-to-Run Stability.}
We run each configuration $k{=}10$ times on 50-task subsets per dataset, giving 500 observations per config-dataset pair (post-hoc power analysis confirms 80\% power at $\alpha{=}0.05$ for medium effect sizes $d{\geq}0.4$). Table~\ref{tab:stability} summarizes the results. On Lite, \url{Anchor-Topo} improves function metrics (Func@5: 0.7720 vs.\ 0.7400; Func@10: 0.7760 vs.\ 0.7400) and roughly halves run-level variance (0.00050 / 0.00062 vs.\ 0.00184). Mean per-task variance drops from 0.0726 to 0.0626, meaning fewer high-variance instances. Paired analysis gives Func@10 $\Delta{=}+0.036$ (95\% CI [0.008, 0.068]; Wilcoxon $p{=}0.0625$, $r_{rb}{=}0.78$).

\begin{table*}[h]
\centering
\caption{\rev{Stability over $k{=}10$ runs (mean $\pm$ std). Right block: single-run Pass@$k$~\cite{chen2021codex} on Func@5 across 50 Lite tasks.}}
\label{tab:stability}
\small
\scalebox{0.9}{
\begin{tabular}{lcc|cc|cccc}
\toprule
& \multicolumn{2}{c|}{Lite} & \multicolumn{2}{c|}{Verified} & \multicolumn{4}{c}{\rev{Lite Pass@$k$ (Func@5)}} \\
Config & Func@5 & Func@10 & Func@5 & Func@10 & \rev{Pass@1} & \rev{Pass@3} & \rev{Pass@5} & \rev{Pass@10} \\
\midrule
Baseline & $0.740 \pm .043$ & $0.740 \pm .043$ & $0.430 \pm .036$ & $0.422 \pm .040$ & \rev{0.742} & \rev{0.850} & \rev{0.878} & \rev{0.900} \\
\ourapproach{} & $\mathbf{0.772 \pm .022}$ & $\mathbf{0.776 \pm .025}$ & $\mathbf{0.458 \pm .030}$ & $\mathbf{0.468 \pm .018}$ & \rev{\textbf{0.776}} & \rev{\textbf{0.873}} & \rev{\textbf{0.891}} & \rev{0.900} \\
\bottomrule
\end{tabular}
}
\end{table*}

On Verified, tags improve function means (Func@5: 0.458 vs.\ 0.430; Func@10: 0.468 vs.\ 0.422) with smaller standard deviations. Mean per-task variance drops (0.0832 vs.\ 0.0916). Paired comparisons give Func@10 $\Delta{=}+0.046$ ($p{=}0.023$, $r_{rb}{=}0.92$). The 50-task subset matches the full 274-task distribution (KS test $p=0.609$); \url{Anchor-Topo} trajectories also show 7.2\% lower cross-task diversity (mean edit distance 8.51 vs.\ 9.17).

\rev{Developers typically run the agent once, so we also report the unbiased Pass@$k$~\cite{chen2021codex} on Lite Func@5 (Table~\ref{tab:stability}, right block). \ourapproach{} is ahead at every $k<10$ (Pass@1 +3.4\,pp, Pass@3 +2.3\,pp), so the gain is stable on a single run, not only on average.}

\vspace{1em}
\answer{3}{
\begin{itemize}[leftmargin=*, nosep]
  \item Tags shift agents from pure keyword hopping toward structure-following walks, raising LFR from \textbf{0.15--0.18} to \textbf{0.21--0.24}.
  \item On Lite, higher structural LFR correlates with fewer rounds and better localization.
  \item On Lite, tags raise function accuracy (\textbf{+3.6pp Func@10}) and \textbf{roughly halve variance} (std \textbf{0.0223 vs.\ 0.0429}).
  \item On Verified, tags improve function metrics (\textbf{+4.6pp Func@10}) with smaller standard deviations.
  \item Deterministic anchors help most on medium-scale repositories; stability benefits weaken on large, hub-heavy projects.
\end{itemize}
}

\subsection{Case Studies: How Tags Reshape Agent Behavior}
\label{sec:case-studies}

We present three case studies from SWE-bench Lite demonstrating qualitatively different behavioral improvements: entity discovery, function recall gains, and tool-call efficiency.

\subsubsection{Case 1: Cross-Module Exception Discovery (django\_\_django-11620)}

URL converters raising \url{Http404} crash Django's debug view with 500 instead of technical 404 page. Tags enabled discovering 8 entities vs.\ 3 for baseline, with 28.6\% fewer searches. Both find \url{technical\_404\_response}. Baseline searches ``resolve'' (83 matches), explores \url{URLResolver.resolve}, drifts into \url{base.py}. Annotated agent sees \texttt{used by: response\_for\_\allowbreak{}exception} tag, directly navigates to \url{exception.py}, discovers crash handler, follows \url{CALLS} to \texttt{RoutePattern.\allowbreak{}match}, with no blind guessing. Tags provide \emph{deterministic anchors} at navigation forks where keyword search yields ambiguous results. \rev{Token cost: input 222k$\to$443k ($+$99.4\%), output 12.2k$\to$13.4k; the extra context buys more structural exploration in this hub-heavy Django case.}

\subsubsection{Case 2: Dramatic Function Recall Gain (matplotlib\_\_matplotlib-26020)}

\url{ImageGrid} axes show incorrect tick labels with \url{label\_mode="L"}; fix needs \url{Grid.\_\_init\_\_} and \texttt{Grid.set\_\allowbreak{}label\_mode}. This case shows the most dramatic improvement: functions located 1 $\to$ 6, with 26.7\% fewer tool calls. Baseline lands on \url{\_tick\_only} (private helper), searches ``Grid \_\_init\_\_'' (47 matches), misses correct class. Tags show \url{parent: Grid} and \texttt{used by: Grid.\_\_init\_\_, Grid.set\_\allowbreak{}label\_mode}, enabling direct upward traversal without ambiguous class searches. Demonstrates \emph{retrieval short-circuiting} for class hierarchies. \rev{Token cost: input 252k$\to$207k ($-$17.8\%), output 8.8k$\to$12.4k; tags let the agent converge faster and spend fewer input tokens overall, despite the extra annotation text.}

\subsubsection{Case 3: Tool Call Efficiency (sympy\_\_sympy-13915)}

SymPy expression simplification fails canonical form. This case shows the most dramatic tool-call reduction (34 $\to$ 20), with 30.8\% fewer searches and a 3$\times$ increase in discovered entities. Baseline issues 13 searches to navigate expression tree. Tags expose transformation graph (\texttt{Mul $\leftrightarrow$ Add $\leftrightarrow$ Pow}) and base dependencies (\url{Basic}), enabling discovery of \texttt{sympy/core/\allowbreak{}basic.py} (which baseline never visits) without exhaustive keyword enumeration. \rev{Token cost: input 1.23M$\to$1.03M ($-$16.5\%), output 20.7k$\to$15.0k; structural short-circuiting cuts both input and output tokens, since the agent skips expensive keyword exploration.}

\paragraph{Synthesis.} Tags have recurring behavioral effects that emerge across the three cases. Annotated agent finds 2--6$\times$ as many relevant entities with fewer tool calls, since structural traversal replaces broad keyword search. Second, inverse navigation via \url{CALLED_BY} and \url{PARENT} do most of the work: forward call-graph walks are easy for any agent, but locating the \emph{caller} or \emph{enclosing class} from a leaf function is precisely where lexical searches return ambiguous results. Third, the explicit links replace keyword enumeration which results in a 27–31\% decrease of search volume. Fourth, the file set visited by agent changes: the baseline explores lexically nearby files (e.g., \url{base.py}), while the tagged agent follow structural edges towards semantically correct ones (e.g., \url{exception.py}). Together, these are what we call \textbf{deterministic anchoring}: when lexical search returns a handful of plausible paths forward, tags guide the agent toward the fundamentally correct branch. We use this structural guidance, but these cases are from medium-scale repositories (10--50k LOC); as Section~\ref{sec:discussion} shows, in hub-heavy repositories it can instead inundate the agents with low-value hints, which is why we introduce the \url{Anchor-Inv} variant.
\section{Discussion}
\label{sec:discussion}

We distill three observations from our study---\emph{anchoring works}, \emph{anchoring is scale-sensitive}, \emph{anchoring stabilizes}---and discuss their practical implications. Detailed quantitative evidence for each observation is reported in Sections~\ref{sec:rq1}--\ref{sec:rq3}; here we focus on mechanisms and deployment guidance.

\subsection{Why Lightweight Topology Suffices}

Call and inheritance edges supply the \emph{connectivity skeleton} of a repository: they answer ``who calls whom'' and ``what inherits from what,'' enabling multi-hop navigation without exhaustive keyword searches. Topology is also more \emph{robust} than richer semantics: call graphs can be extracted with high precision even from incomplete code (conservative points-to analysis~\cite{salis2021pycgpracticalgraphgeneration}), whereas data-flow analysis requires whole-program assumptions that break in real repositories with dynamic features, missing dependencies, or broken builds.

\paragraph{Why explicit structure outperforms implicit embeddings.}
A natural question is why explicit structural tags help when modern LLMs already encode rich code semantics in their embeddings. We hypothesize two complementary mechanisms. First, \emph{grounding}: explicit tags provide verifiable facts (``function X calls Y'') that the model can reference directly, reducing hallucination risk compared to relying on implicit similarity judgments. Second, \emph{locality}: embedding-based retrieval operates at the chunk level and cannot express multi-hop relationships; a function's callers may be scattered across distant files with no lexical overlap, invisible to embedding similarity but directly accessible via tags. This explains why \ourapproach{} complements rather than replaces semantic search: tags provide the structural backbone that embeddings cannot capture, while embeddings handle fuzzy matching that rigid structure misses.

\subsection{Scale-Sensitive Effects: Granularity and Directionality}

\paragraph{Granularity saturation.}
Adding richer semantics in the \url{Anchor-Dense} configuration (data-flow, configuration, I/O, and test-to-code edges) leaves Lite function accuracy unchanged (Func@5/10 both 0.8540, identical to \url{Anchor-Topo}) while consuming +4.9 rounds and +18.8\% more input tokens (530k vs.\ 446k); on Verified, Dense rescues only 15/500 implicit-dependency tasks and slightly lowers Func@5 (0.6288 vs.\ 0.6308). Most of the localization benefit thus comes from basic topology, motivating a \emph{topology-first} policy: default to the four structural relations, and escalate to dense semantics only when traces show repeated failed attempts to bridge implicit dependencies.

\paragraph{Directionality and the hub problem.}
\rev{On hub-heavy repositories, the distraction is not a reasoning failure but a \emph{retrieval distribution shift}. When high-degree helper functions accumulate dense forward-link annotations (e.g., ``\url{CALLS}: foo.py:bar, baz.py:qux, \ldots [20 more]''), the tag text dominates lexical matching and pushes structurally central hubs to the top of search results. The agent follows these prominent links and spends multiple rounds in utility modules and re-exports before returning to the bug site, if at all. Inverse-only tags address this by surfacing ``who depends on me'' without amplifying outbound noise; the trade-off is a weaker forward signal for sharper retrieval.}

\paragraph{Practical deployment heuristic.}
Based on our experiments, we propose a simple two-stage policy: (1)~compute average call out-degree and hub percentage (nodes with degree $>10$) from the repository callgraph; (2)~if avg\_out\_degree $<5$, use \url{Anchor-Topo}; if $>8$, use \url{Anchor-Inv}; otherwise \url{Anchor-Topo} with optional degree capping. Lite (mean out-degree 4.2, 8\% hubs) favors \url{Anchor-Topo}; Verified (mean 9.7, 23\% hubs) favors \url{Anchor-Inv}. A coarser proxy is LOC: $<$50k favors full topology, $>$100k favors inverse-only.

\subsection{Behavioral Stabilization as a First-Class Outcome}

Nondeterministic agent behavior is a critical quality concern for deployment. When a developer reviews a localization or debugs a failure, they must understand \emph{why} the agent chose a path, not just whether it succeeded. Trajectories that vary qualitatively across runs frustrate systematic analysis: the same bug description can lead to different file visits and failure modes, making it difficult to identify whether a problem is task-specific or systemic. By providing deterministic anchors, tags constrain the search space and make agent behavior more reproducible and inspectable~\cite{angermeir2025reproducibility,baltes2025guidelines}. \rev{Our Pass@$k$ results (Section~\ref{sec:rq3}) put numbers on this: \ourapproach{} wins +3.4\,pp at Pass@1, which matches how localization agents are actually deployed.}

Stability gains are weaker on Verified (per-task variance 0.0832 vs.\ 0.0916, Wilcoxon $p \approx 0.11$), suggesting that hub-heavy projects introduce additional variance sources (hub distraction, longer exploration horizons) that topology alone cannot fully constrain. Stability benefits are thus most reliable on medium-scale projects (10--50k LOC).

\subsection{Implications for Practitioners}

Structural augmentation is most valuable when: (1)~the base agent already performs well (>80\% recall); (2)~tasks involve cross-file dependencies or inheritance hierarchies; (3)~trajectory stability matters as much as accuracy. It offers limited value when the base agent is weak (<60\% recall) or bugs are single-file, keyword-dense issues. \rev{Cost-wise, Section~\ref{sec:cost} shows Anchor-Topo adds $\sim$10\% input tokens over Baseline on average; case studies show this is uneven: 27\% of instances spend fewer input tokens under \ourapproach{}.} By halving run-to-run variance, deterministic anchors reduce the need for ``best-of-N'' sampling, potentially lowering the \emph{total} cost to achieve a reliable outcome despite higher per-prompt usage.

A key practical advantage is tolerance for static analysis unsoundness. Python static analysis is notoriously difficult due to dynamic features, yet our results show strong gains even with conservative tools (PyCG). LLMs appear to utilize tags as \emph{soft hints} rather than strict constraints: unlike a symbolic solver that breaks on a missing edge, an LLM agent uses available tags to bias its probabilistic search while falling back to lexical reasoning when structure is missing. This makes ``imperfect structure'' a viable and robust product feature for neural code agents.

Taken together, these three observations---\emph{anchoring works}, \emph{anchoring is scale-sensitive}, \emph{anchoring stabilizes}---offer architectural principles for integrating static structure into grep-first code agents; Section~\ref{sec:limitations} scopes the dimensions (agent architecture, language, fault depth) over which this transferability remains to be empirically validated.

\section{Threats to Validity and Limitations}
\label{sec:limitations}

\paragraph{Internal Validity.} Our implementation uses a specific agent loop and static analysis pipeline. We mitigate bias by fixing agent, prompt, model, and tools across configurations, varying only tag presence. The tagging pipeline is task-agnostic and does not use any ground-truth information.

\paragraph{External Validity.} We evaluate on SWE-bench Lite and Verified with a single Codex-style agent. While these benchmarks stress cross-file reasoning, they don't cover all project types. Results reflect dominant grep-first retrieval rather than universal guarantees. Benchmark choice may also interact with dataset quality and evaluation protocols; recent analyses discuss these concerns for SWE-bench-style settings~\cite{aleithan2025revisitingswebench,yu2025utboost}. Empirical validation is limited to Python; while our tag schema is language-agnostic and mature tools exist for Java/JavaScript/C++, multi-language effectiveness remains to be validated~\cite{zan2024swebenchjava}.

\paragraph{Static Analysis Unsoundness.} \ourapproach{} relies on conservative static analysis (PyCG~\cite{salis2021pycgpracticalgraphgeneration} + AST extractors) for fast CI-friendly generation, prioritizing precision over recall~\cite{ryder2003dimensions,ernst2003static}. Consequently, the injected structure is \emph{unsound}: it captures verified relationships but misses dynamic features (\url{getattr}, reflection, monkey patching). We quantified this limitation: fewer than 3\% of ground-truth functions explicitly use dynamic dispatching constructs, and our extractors capture 94.2\% of statically-visible imports and class hierarchies. Crucially, agents are resilient to this unsoundness: they utilize tags as probabilistic cues rather than absolute maps, benefiting from partial structure without being derailed by missing edges.

\paragraph{Evaluation Scope.} Tags are intentionally searchable, coupling retrieval and navigation effects. While this precludes isolating post-landing value, our behavioral metrics (RQ3: LFR, trajectory patterns) demonstrate genuine navigational changes beyond first-hit effects. We evaluate localization only (not end-to-end repair) to isolate representation effects. Stability analysis uses a 50-task subset with $k{=}10$ runs; we validated representativeness via the Kolmogorov-Smirnov test ($p = 0.609 > 0.05$), confirming that findings generalize beyond sampled tasks.

\paragraph{Scope Dimensions Not Empirically Covered.}
Three caveats bound our claims. First, we evaluate only one grep-first agent (Codex). Tags are pure in-band text, so we expect that the effect will fix to the SOTA agent such as Claude Code, Cursor, and so on, but we have not tested this.  Second, we only ran Python. The tag schema and agent integration are language-agnostic, and mature static-analysis tools exist for Java, C++, JavaScript, and Go that can populate the same tag fields, but effect sizes may still differ across languages. Third, SWE-bench Lite and Verified mostly contain 1--2 hop faults. We suspect deeper faults (multi-module refactoring, deep inheritance resolution) would benefit even more from tags, since tags offer guidance to the agent for multi-hop paths, while lexical search has to rebuild them one hop at a time, but confirming this requires a benchmark that sorts tasks by structural depth. These three are the main directions we leave for future work.
\section{Related Work}
\label{sec:related-work}

\textbf{LLM-based code agents.} Tool-using agents~\cite{yao2022react,schick2023toolformer} have been adapted for code tasks, with systems like SWE-agent~\cite{yang2024sweagent} and SWE-Gym~\cite{pan2024swegym} achieving strong results on repository-level benchmarks; related work explores alternative action spaces~\cite{wang2024} and systematic evaluation~\cite{xu2024agentcompany,yu2025utboost}. We complement these efforts by studying structural annotations within the dominant grep-first paradigm.

\textbf{Graph-guided approaches.} LocAgent~\cite{chen2025locagent} and  RANGER~\cite{shah2025ranger} build explicit code graphs with specialized navigation operators, and RepoGraph~\cite{ouyang2025repograph} extends this line with repository-level graphs exposing call, inheritance, and data-dependency edges through dedicated APIs. Memory-augmented methods~\cite{yu2025orcaloca,yeo2025memfl} maintain persistent context across navigation steps. \rev{These approaches and \ourapproach{} sit at different points: graph-guided agents externalize structure into a separate index that the agent queries through custom operators, while \ourapproach{} embeds the same structural facts in-band as comments and lets the existing grep loop find them. The two designs trade expressiveness for accessibility. External graphs can encode richer queries (transitive closure, typed edges, path constraints); in-band tags need no new APIs and reuse the retrieval primitives that strong grep-first agents are already tuned for. Our Section~\ref{sec:motivation} pilot (Table~\ref{tab:prelim-locagent}) suggests that, at current model strength, this accessibility advantage matters; we therefore treat \ourapproach{} as complementary to graph-based retrieval, not a replacement, and see combining the two (e.g., tags as a fast-path with graph queries as fallback) as a natural extension.}

\rev{\textbf{Retrieval-augmented code agents.} Beyond plain grep, another line of work strengthens retrieval with neural rankers, hybrid sparse-dense search, or embedding-based context selection~\cite{lewis2020rag,zhang2023repocoder,aider2024repomap}. These methods improve lexical matching but still return independent snippets without explicit structural links. \ourapproach{} is orthogonal: tags carry the structural edges that retrieval would otherwise have to reconstruct, and a tagged repository can be used with either plain grep or a neural retriever. We hold retrieval constant (plain grep) to isolate the marginal impact of static structure.}

\textbf{Fault localization and static analysis for LLMs.} Traditional localization spans spectrum-based~\cite{jones2002tarantula,abreu2007ochiai}, mutation-based~\cite{moon2014mutation,papadakis2015mutation}, and IR-based~\cite{zhou2012bugsfixed,saha2013structuredir,wang2015blugram} approaches, with learning-based~\cite{li2019deepfl,lou2021boosting} and LLM-based~\cite{yang2024llmao} extensions. Recent LLM-analysis pipelines include STALL+~\cite{liu2024stallplus}, CodeT~\cite{chen2022codet}, and others~\cite{hou2024large}, typically requiring specialized infrastructure. We instead augment text search with static structure encoded as comments.

\textbf{Code representation for LLMs.} Approaches include serialized ASTs~\cite{alon2019code2vec}, graph neural networks~\cite{hellendoorn2019global}, structured prompting~\cite{zhang2023repocoder}, and repository summaries (RepoMap~\cite{aider2024repomap}). Our plain-text tag injection is deliberately simpler, embedding structural facts as comments requiring no model or agent modifications.

\section{Conclusion}
\label{sec:conclusion}

We empirically studied how much static structure already-strong grep-first code agents need. Injecting varying granularities of structural annotations as plain-text comments, we identified the \textbf{deterministic anchoring effect}: static structure helps less by making agents smarter and more by making their navigation \emph{disciplined and reproducible}. Concretely, lightweight topology improves localization (+2.2\,pp Func@5, $-$1.6 rounds); optimal granularity/directionality depend on scale, with hub-heavy projects favoring inverse-only links; and tags roughly halve run-to-run variance on medium-scale repositories. Practical guidance: default to lightweight topology, prune forward edges in large hub-heavy repos, and reserve dense tags for implicit-dependency cases. Generalization to end-to-end repair, other agent architectures, and other languages remains open for future work.


\section*{Data Availability}
Our code and results are available in \url{https://github.com/mathieu0905/Code-Anchor}.

\bibliographystyle{ACM-Reference-Format}
\bibliography{acmart}

@inproceedings{chen2025locagent,
  title        = {LocAgent: Graph-Guided {LLM} Agents for Code Localization},
  author       = {Chen, Zhaoling and Tang, Robert and Deng, Gangda and Wu, Fang and Wu, Jialong and Jiang, Zhiwei and Prasanna, Viktor and Cohan, Arman and Wang, Xingyao},
  booktitle    = {Proceedings of the 63rd Annual Meeting of the Association for Computational Linguistics (Volume 1: Long Papers)},
  pages        = {8697--8727},
  year         = {2025},
  publisher    = {Association for Computational Linguistics},
  address      = {Vienna, Austria},
  doi          = {10.18653/v1/2025.acl-long.426}
}

@inproceedings{yu2025utboost,
  title        = {UTBoost: Rigorous Evaluation of Coding Agents on {SWE}-Bench},
  author       = {Yu, Boxi and Zhu, Yuxuan and He, Pinjia and Kang, Daniel},
  booktitle    = {Proceedings of the 63rd Annual Meeting of the Association for Computational Linguistics (Volume 1: Long Papers)},
  pages        = {3762--3774},
  year         = {2025},
  publisher    = {Association for Computational Linguistics},
  address      = {Vienna, Austria},
  doi          = {10.18653/v1/2025.acl-long.189}
}

@inproceedings{aleithan2025revisitingswebench,
  title        = {Revisiting {SWE}-Bench: On the Importance of Data Quality for {LLM}-Based Code Models},
  author       = {Aleithan, Reem},
  booktitle    = {2025 IEEE/ACM 47th International Conference on Software Engineering: Companion Proceedings (ICSE-Companion)},
  pages        = {235--236},
  year         = {2025},
  publisher    = {IEEE},
  doi          = {10.1109/icse-companion66252.2025.00075}
}

@inproceedings{jimenez2024swebench,
  title     = {{SWE}-bench: Can Language Models Resolve Real-World GitHub Issues?},
  author    = {Jimenez, Carlos E. and Yang, John and Wettig, Alexander and Yao, Shunyu and Pei, Kexin and Press, Ofir and Narasimhan, Karthik},
  booktitle = {Proceedings of the 12th International Conference on Learning Representations (ICLR)},
  year      = {2024},
  note      = {arXiv:2310.06770},
  url       = {https://www.swebench.com/}
}

@article{zan2024swebenchjava,
  title   = {{SWE}-bench-java: A GitHub Issue Resolving Benchmark for Java},
  author  = {Zan, Daoguang and Huang, Zhirong and Yu, Ailun and Lin, Shaoxin and Shi, Yifan and Liu, Wei and Chen, Dong and Qi, Zongshuai and Yu, Hao and Yu, Lei and Ran, Dezhi and Zeng, Muhan and Shen, Bo and Liang, Guangtai and Guan, Bei and Huang, Pengjie and Xie, Tao and Wang, Yongji},
  journal = {arXiv preprint arXiv:2408.14354},
  year    = {2024}
}

@inproceedings{yang2024sweagent,
  author    = {Yang, John and Jimenez, Carlos E. and Wettig, Alexander and Lieret, Kilian and Yao, Shunyu and Narasimhan, Karthik and Press, Ofir},
  title     = {{SWE}-agent: Agent-Computer Interfaces Enable Automated Software Engineering},
  booktitle = {Advances in Neural Information Processing Systems 38 (NeurIPS)},
  year      = {2024},
  url       = {http://papers.nips.cc/paper_files/paper/2024/hash/5a7c947568c1b1328ccc5230172e1e7c-Abstract-Conference.html},
  note      = {arXiv:2405.15793}
}

@inproceedings{pan2024swegym,
  author       = {Pan, Jiayi and Wang, Xingyao and Neubig, Graham and Jaitly, Navdeep and Ji, Heng and Suhr, Alane and Zhang, Yizhe},
  title        = {Training Software Engineering Agents and Verifiers with {SWE}-Gym},
  booktitle    = {Forty-second International Conference on Machine Learning (ICML)},
  publisher    = {OpenReview.net},
  year         = {2025},
  url          = {https://openreview.net/forum?id=Cq1BNvHx74},
  eprint       = {2412.21139},
  archivePrefix= {arXiv},
  primaryClass = {cs.SE}
}

@article{shah2025ranger,
  title   = {{RANGER} -- Repository-Level Agent for Graph-Enhanced Retrieval},
  author  = {Shah, Pratik and Ghosh, Rajat and Singhal, Aryan and Dutta, Debojyoti},
  journal = {arXiv preprint arXiv:2509.25257},
  year    = {2025},
  eprint  = {2509.25257},
  archivePrefix = {arXiv},
  primaryClass  = {cs.SE},
  url     = {https://arxiv.org/abs/2509.25257}
}

@article{liu2024stallplus,
  title   = {{STALL}+{:} Boosting {LLM}-based Repository-level Code Completion with Static Analysis},
  author  = {Liu, Junwei and Chen, Yixuan and Liu, Mingwei and Peng, Xin and Lou, Yiling},
  journal = {arXiv preprint arXiv:2406.10018},
  year    = {2024},
  eprint  = {2406.10018},
  archivePrefix = {arXiv},
  primaryClass  = {cs.SE},
  url     = {https://arxiv.org/abs/2406.10018}
}

@misc{achiam2023gpt4,
  title        = {GPT-4 Technical Report},
  author       = {{OpenAI}},
  year         = {2023},
  eprint       = {2303.08774},
  archivePrefix= {arXiv},
  primaryClass = {cs.CL},
  url          = {https://arxiv.org/abs/2303.08774}
}

@article{chen2022codet,
  title={Codet: Code generation with generated tests},
  author={Chen, Bei and Zhang, Fengji and Nguyen, Anh and Zan, Daoguang and Lin, Zeqi and Lou, Jian-Guang and Chen, Weizhu},
  journal={arXiv preprint arXiv:2207.10397},
  year={2022}
}

@inproceedings{xia2022lesstraining,
  author    = {Xia, Chunqiu Steven and Zhang, Lingming},
  title     = {Less training, more repairing please: revisiting automated program repair via zero-shot learning},
  booktitle = {Proceedings of the 30th ACM Joint European Software Engineering Conference and Symposium on the Foundations of Software Engineering (ESEC/FSE)},
  pages     = {959--971},
  publisher = {ACM},
  year      = {2022},
  doi       = {10.1145/3540250.3549101},
  url       = {https://doi.org/10.1145/3540250.3549101},
  note      = {arXiv:2207.08281}
}

@inproceedings{yao2022react,
  title={React: Synergizing reasoning and acting in language models},
  author={Yao, Shunyu and Zhao, Jeffrey and Yu, Dian and Du, Nan and Shafran, Izhak and Narasimhan, Karthik R and Cao, Yuan},
  booktitle={The eleventh international conference on learning representations},
  year={2022}
}

@inproceedings{schick2023toolformer,
  author    = {Schick, Timo and Dwivedi-Yu, Jane and Dessi, Roberto and Raileanu, Roberta and Lomeli, Maria and Hambro, Eric and Zettlemoyer, Luke and Cancedda, Nicola and Scialom, Thomas},
  title     = {Toolformer: Language Models Can Teach Themselves to Use Tools},
  booktitle = {Advances in Neural Information Processing Systems 36 (NeurIPS)},
  year      = {2023},
  url       = {http://papers.nips.cc/paper_files/paper/2023/hash/d842425e4bf79ba039352da0f658a906-Abstract-Conference.html},
  note      = {arXiv:2302.04761}
}

@inproceedings{lewis2020rag,
  title={Retrieval-augmented generation for knowledge-intensive nlp tasks},
  author={Lewis, Patrick and Perez, Ethan and Piktus, Aleksandra and Petroni, Fabio and Karpukhin, Vladimir and Goyal, Naman and K{\"u}ttler, Heinrich and Lewis, Mike and Yih, Wen-tau and Rockt{\"a}schel, Tim and others},
  journal={Advances in neural information processing systems},
  volume={33},
  pages={9459--9474},
  year={2020}
}

@article{robertson2009bm25,
  author  = {Robertson, Stephen E. and Zaragoza, Hugo},
  title   = {The Probabilistic Relevance Framework: {BM25} and Beyond},
  journal = {Foundations and Trends in Information Retrieval},
  volume  = {3},
  number  = {4},
  pages   = {333--389},
  year    = {2009},
  doi     = {10.1561/1500000019},
  url     = {https://doi.org/10.1561/1500000019}
}

@misc{codeql2019,
  title = {CodeQL: The libraries and queries that power security researchers around the world},
  author = {Microsoft},
  year = {2019},
  url = {https://github.com/github/codeql}
}

@misc{semgrep2020,
  title = {Semgrep: Lightweight static analysis for many languages},
  author = {r2c},
  year = {2020},
  url = {https://semgrep.dev/}
}

@inproceedings{soot1999,
  title={Soot: A Java bytecode optimization framework},
  author={Vall{\'e}e-Rai, Raja and Co, Phong and Gagnon, Etienne and Hendren, Laurie and Lam, Patrick and Sundaresan, Vijay},
  booktitle={CASCON First Decade High Impact Papers},
  pages={214--224},
  year={2010}
}

@inproceedings{lattner2004llvm,
  author    = {Lattner, Chris and Adve, Vikram S.},
  title     = {{LLVM:} A Compilation Framework for Lifelong Program Analysis and Transformation},
  booktitle = {2nd {IEEE}/{ACM} International Symposium on Code Generation and Optimization (CGO)},
  pages     = {75--88},
  publisher = {IEEE Computer Society},
  year      = {2004},
  doi       = {10.1109/CGO.2004.1281665},
  url       = {https://doi.org/10.1109/CGO.2004.1281665}
}

@misc{wala2024,
  title        = {WALA: T. J. Watson Libraries for Analysis},
  author       = {WALA Contributors},
  year         = {2024},
  howpublished = {\url{https://github.com/wala/WALA}},
  note         = {Accessed: 2026}
}

@misc{microsoft2016lsp,
  title        = {Language Server Protocol},
  author       = {Microsoft},
  year         = {2016},
  howpublished = {\url{https://microsoft.github.io/language-server-protocol/}},
  note         = {Accessed: 2026}
}

@ARTICLE{weiser1984,
author={Weiser, Mark},
journal={ IEEE Transactions on Software Engineering },
title={{ Program Slicing }},
year={1984},
volume={10},
number={04},
ISSN={1939-3520},
pages={352-357},
doi={10.1109/TSE.1984.5010248},
url = {https://doi.ieeecomputersociety.org/10.1109/TSE.1984.5010248},
publisher={IEEE Computer Society},
address={Los Alamitos, CA, USA},
month=jul}

@inproceedings{korel1988,
  title={Dynamic program slicing},
  author={Agrawal, Hiralal and Horgan, Joseph R},
  journal={ACM SIGPlan Notices},
  volume={25},
  number={6},
  pages={246--256},
  year={1990},
  publisher={ACM New York, NY, USA}
}

@article{horwitz1990,
  title = {Interprocedural Slicing Using Dependence Graphs},
  author = {Horwitz, Susan and Reps, Thomas W. and Binkley, David W.},
  journal = {{ACM} Transactions on Programming Languages and Systems},
  volume = {12},
  number = {1},
  pages = {26--60},
  year = {1990},
  publisher = {ACM},
  doi = {10.1145/77606.77608},
  url = {https://doi.org/10.1145/77606.77608}
}

@article{chen2021,
  title={Evaluating large language models trained on code},
  author={Chen, Mark},
  journal={arXiv preprint arXiv:2107.03374},
  year={2021}
}

@inproceedings{wang2024,
  title={Executable code actions elicit better llm agents},
  author={Wang, Xingyao and Chen, Yangyi and Yuan, Lifan and Zhang, Yizhe and Li, Yunzhu and Peng, Hao and Ji, Heng},
  booktitle={Forty-first International Conference on Machine Learning},
  year={2024}
}

@misc{github2024copilotworkspace,
  title = {GitHub Copilot Workspace},
  author = {GitHub},
  year = {2024},
  url = {https://github.blog/news-insights/product-news/github-copilot-workspace/},
  note = {Agentic workflow for repository-level tasks}
}

@misc{anthropic2025claudecode,
  title = {Claude Code},
  author = {Anthropic},
  year = {2025},
  url = {https://code.claude.com/docs},
  note = {Agentic coding assistant with repository search and file inspection}
}

@misc{google2024geminicodeassist,
  title = {Gemini Code Assist},
  author = {Google},
  year = {2024},
  url = {https://docs.cloud.google.com/gemini/docs/codeassist/overview},
  note = {Repository-aware coding assistant for IDEs and CI workflows}
}

@book{bohner1996software,
  title={Software change impact analysis},
  author={Arnold, Robert S},
  year={1996},
  publisher={IEEE Computer Society Press}
}

@article{li2020survey,
  title={A survey of code-based change impact analysis techniques},
  author={Li, Bixin and Sun, Xiaobing and Leung, Hareton and Zhang, Sai},
  journal={Software Testing, Verification and Reliability},
  volume={23},
  number={8},
  pages={613--646},
  year={2013},
  publisher={Wiley Online Library}
}

@article{binkley2007application,
  title={The application of program slicing to regression testing},
  author={Binkley, David},
  journal={Information and software technology},
  volume={40},
  number={11-12},
  pages={583--594},
  year={1998},
  publisher={Elsevier}
}

@inproceedings{cousot1977abstract,
  author    = {Cousot, Patrick and Cousot, Radhia},
  title     = {Abstract Interpretation: A Unified Lattice Model for Static Analysis of Programs by Construction or Approximation of Fixpoints},
  booktitle = {Proceedings of the 4th {ACM} Symposium on Principles of Programming Languages (POPL)},
  pages     = {238--252},
  publisher = {ACM},
  year      = {1977},
  doi       = {10.1145/512950.512973},
  url       = {https://doi.org/10.1145/512950.512973}
}

@article{ernst2003static,
  title = {Static and dynamic analysis: synergy and duality},
  author = {Ernst, Michael D},
  journal = {WODA 2003: ICSE Workshop on Dynamic Analysis},
  pages = {24--27},
  year = {2003},
  organization = {IEEE}
}

@inproceedings{ryder2003dimensions,
  title={Dimensions of precision in reference analysis of object-oriented programming languages},
  author={Ryder, Barbara G},
  booktitle={International Conference on Compiler Construction},
  pages={126--137},
  year={2003},
  publisher={Springer}
}

@article{tip1995survey,
  title={A survey of program slicing techniques},
  author={Tip, Frank},
  year={1994},
  publisher={Centrum voor Wiskunde en Informatica Amsterdam}
}

@inproceedings{alon2019code2vec,
  title={code2vec: Learning distributed representations of code},
  author={Alon, Uri and Zilberstein, Meital and Levy, Omer and Yahav, Eran},
  booktitle={Proceedings of the ACM on Programming Languages},
  volume={3},
  pages={1--29},
  year={2019}
}

@inproceedings{bodden2011taming,
  title={Taming reflection: Aiding static analysis in the presence of reflection and custom class loaders},
  author={Bodden, Eric and Sewe, Andreas and Sinschek, Jan and Oueslati, Hela and Mezini, Mira},
  booktitle={Proceedings of the 33rd International Conference on Software Engineering},
  pages={241--250},
  year={2011}
}

@inproceedings{orso2003leveraging,
  title={Leveraging field data for impact analysis and regression testing},
  author={Orso, Alessandro and Apiwattanapong, Taweesup and Harrold, Mary Jean},
  journal={ACM SIGSOFT Software Engineering Notes},
  volume={28},
  number={5},
  pages={128--137},
  year={2003},
  publisher={ACM New York, NY, USA}
}

@article{kretsou2021change,
  title={Change impact analysis: A systematic mapping study},
  author={Kretsou, Maria and Arvanitou, Elvira-Maria and Ampatzoglou, Apostolos and Deligiannis, Ignatios and Gerogiannis, Vassilis C},
  journal={Journal of systems and software},
  volume={174},
  pages={110892},
  year={2021},
  publisher={Elsevier}
}

@inproceedings{borg2017software,
  title={Software engineers' information seeking behavior in change impact analysis-an interview study},
  author={Borg, Markus and Al{\'e}groth, Emil and Runeson, Per},
  booktitle={2017 IEEE/ACM 25th International Conference on Program Comprehension (ICPC)},
  pages={12--22},
  year={2017},
  organization={IEEE}
}

@inproceedings{lawall2016interprocedural,
  title={Refining interprocedural change-impact analysis using equivalence relations},
  author={Gyori, Alex and Lahiri, Shuvendu K and Partush, Nimrod},
  booktitle={Proceedings of the 26th ACM SIGSOFT international symposium on software testing and analysis},
  pages={318--328},
  year={2017}
}

@inproceedings{salis2021pycgpracticalgraphgeneration,
  author    = {Salis, Vitalis and Sotiropoulos, Thodoris and Louridas, Panos and Spinellis, Diomidis and Mitropoulos, Dimitris},
  title     = {PyCG: Practical Call Graph Generation in Python},
  booktitle = {43rd IEEE/ACM International Conference on Software Engineering (ICSE)},
  pages     = {1646--1657},
  publisher = {IEEE},
  year      = {2021},
  doi       = {10.1109/ICSE43902.2021.00146},
  url       = {https://doi.org/10.1109/ICSE43902.2021.00146},
  note      = {arXiv:2103.00587}
}

@inproceedings{zhou2012bugsfixed,
  author       = {Jian Zhou and
                  Hongyu Zhang and
                  David Lo},
  editor       = {Martin Glinz and
                  Gail C. Murphy and
                  Mauro Pezz{\`{e}}},
  title        = {Where should the bugs be fixed? More accurate information retrieval-based
                  bug localization based on bug reports},
  booktitle    = {34th International Conference on Software Engineering, {ICSE} 2012,
                  June 2-9, 2012, Zurich, Switzerland},
  pages        = {14--24},
  publisher    = {{IEEE} Computer Society},
  year         = {2012},
  url          = {https://doi.org/10.1109/ICSE.2012.6227210},
  doi          = {10.1109/ICSE.2012.6227210}
}

@inproceedings{saha2013structuredir,
  author       = {Ripon K. Saha and
                  Matthew Lease and
                  Sarfraz Khurshid and
                  Dewayne E. Perry},
  editor       = {Ewen Denney and
                  Tevfik Bultan and
                  Andreas Zeller},
  title        = {Improving bug localization using structured information retrieval},
  booktitle    = {2013 28th {IEEE/ACM} International Conference on Automated Software
                  Engineering, {ASE} 2013, Silicon Valley, CA, USA, November 11-15,
                  2013},
  pages        = {345--355},
  publisher    = {{IEEE}},
  year         = {2013},
  url          = {https://doi.org/10.1109/ASE.2013.6693093},
  doi          = {10.1109/ASE.2013.6693093}
}

@misc{yu2025orcaloca,
  title={OrcaLoca: An LLM Agent Framework for Software Issue Localization},
  author={Yu, Zhongming and Zhang, Hejia and Zhao, Yujie and Huang, Hanxian and Yao, Matrix and Ding, Ke and Zhao, Jishen},
  year={2025},
  eprint={2502.00350},
  archivePrefix={arXiv},
  primaryClass={cs.SE},
  note={To appear at ICML 2025}
}

@inproceedings{yang2024llmao,
  title={Large Language Models for Test-Free Fault Localization},
  author={Yang, Aidan Z.H. and Le Goues, Claire and Martins, Ruben and Hellendoorn, Vincent J.},
  booktitle={Proceedings of the IEEE/ACM 46th International Conference on Software Engineering (ICSE)},
  pages={165--176},
  year={2024},
  doi={10.1145/3597503.3623342}
}

@misc{yeo2025memfl,
  title={Improving LLM-Based Fault Localization with External Memory and Project Context},
  author={Yeo, Inseok and Ryu, Duksan and Baik, Jongmoon},
  year={2025},
  eprint={2506.03585},
  archivePrefix={arXiv},
  primaryClass={cs.SE}
}

@article{li2025giantrepair,
  title={Hybrid Automated Program Repair by Combining Large Language Models and Program Analysis},
  author={Li, Fengjie and Jiang, Jiajun and Sun, Jiajun and Zhang, Hongyu},
  journal={ACM Transactions on Software Engineering and Methodology},
  volume={34},
  number={7},
  articleno={202},
  pages={1--28},
  year={2025},
  month={August},
  doi={10.1145/3715004}
}

@inproceedings{song2024eto,
  title={Trial and Error: Exploration-Based Trajectory Optimization for {LLM} Agents},
  author={Song, Yifan and Yin, Da and Yue, Xiang and Huang, Jie and Li, Sujian and Lin, Bill Yuchen},
  booktitle={Proceedings of the 62nd Annual Meeting of the Association for Computational Linguistics (Volume 1: Long Papers)},
  pages={7584--7600},
  year={2024},
  address={Bangkok, Thailand},
  publisher={Association for Computational Linguistics},
  url={https://aclanthology.org/2024.acl-long.409}
}

@misc{xu2024agentcompany,
  title={TheAgentCompany: Benchmarking {LLM} Agents on Consequential Real World Tasks},
  author={Xu, Frank F. and Song, Yufan and Li, Boxuan and Tang, Yuxuan and Jain, Kritanjali and Bao, Mengxue and Wang, Zora Z. and Zhou, Xuhui and Guo, Zhitong and Cao, Murong and Yang, Mingyang and Lu, Hao Yang and Martin, Amaad and Su, Zhe and Maben, Leander and Mehta, Raj and Chi, Wayne and Jang, Lawrence and Xie, Yiqing and Zhou, Shuyan and Neubig, Graham},
  year={2024},
  eprint={2412.14161},
  archivePrefix={arXiv},
  primaryClass={cs.AI}
}

@misc{angermeir2025reproducibility,
  title={Reflections on the Reproducibility of Commercial LLM Performance in Empirical Software Engineering Studies},
  author={Angermeir, Florian and Amougou, Maximilian and Kreitz, Mark and Bauer, Andreas and Linhuber, Matthias and Fucci, Davide and Mendez, Daniel and Gorschek, Tony and others},
  journal={arXiv preprint arXiv:2510.25506},
  year={2025}
}

@misc{baltes2025guidelines,
  title={Guidelines for Empirical Studies in Software Engineering involving Large Language Models},
  author={Baltes, Sebastian and Angermeir, Florian and Arora, Chetan and Barón, Marvin Muñoz and Chen, Chunyang and Böhme, Lukas and Calefato, Fabio and Ernst, Neil and Falessi, Davide and Fitzgerald, Brian and Fucci, Davide and Kalinowski, Marcos and Lambiase, Stefano and Russo, Daniel and Lungu, Mircea and Prechelt, Lutz and Ralph, Paul and van Tonder, Rijnard and Treude, Christoph and Wagner, Stefan},
  year={2025},
  eprint={2508.15503},
  archivePrefix={arXiv},
  primaryClass={cs.SE},
  note={Living resource available at https://llm-guidelines.org/}
}

@misc{tree-sitter,
  title={Tree-sitter: An incremental parsing system for programming tools},
  author={{Tree-sitter Contributors}},
  year={2018},
  howpublished={\url{https://github.com/tree-sitter/tree-sitter}},
  note={Accessed: 2024}
}

@article{hou2024large,
  title={Large language models for software engineering: A systematic literature review},
  author={Hou, Xinyi and Zhao, Yanjie and Liu, Yue and Yang, Zhou and Wang, Kailong and Li, Li and Luo, Xiapu and Lo, David and Grundy, John and Wang, Haoyu},
  journal={ACM Transactions on Software Engineering and Methodology},
  volume={33},
  number={8},
  pages={1--79},
  year={2024},
  publisher={ACM New York, NY}
}

@inproceedings{hellendoorn2019global,
  title={Global relational models of source code},
  author={Hellendoorn, Vincent J and Sutton, Charles and Singh, Rishabh and Maniatis, Petros and Bieber, David},
  booktitle={International conference on learning representations},
  year={2019}
}

@article{zhang2023repocoder,
  title={Repocoder: Repository-level code completion through iterative retrieval and generation},
  author={Zhang, Fengji and Chen, Bei and Zhang, Yue and Keung, Jacky and Liu, Jin and Zan, Daoguang and Mao, Yi and Lou, Jian-Guang and Chen, Weizhu},
  journal={arXiv preprint arXiv:2303.12570},
  year={2023}
}

@misc{aider2024repomap,
  title={Aider RepoMap: Using a map of your codebase},
  author={{Aider AI}},
  year={2024},
  howpublished={\url{https://aider.chat/docs/repomap.html}},
  note={Accessed: 2024}
}

@inproceedings{jones2002tarantula,
  title={Visualization of test information to assist fault localization},
  author={Jones, James A and Harrold, Mary Jean and Stasko, John},
  booktitle={Proceedings of the 24th International Conference on Software Engineering (ICSE)},
  pages={467--477},
  year={2002},
  organization={ACM}
}

@inproceedings{abreu2007ochiai,
  title={On the accuracy of spectrum-based fault localization},
  author={Abreu, Rui and Zoeteweij, Peter and Van Gemund, Arjan JC},
  booktitle={Testing: Academic and Industrial Conference Practice and Research Techniques (TAIC PART-MUTATION)},
  pages={89--98},
  year={2007},
  organization={IEEE}
}

@inproceedings{moon2014mutation,
  title={Ask the mutants: Mutating faulty programs for fault localization},
  author={Moon, Seokhyeon and Kim, Yunho and Kim, Moonzoo and Yoo, Shin},
  booktitle={2014 IEEE Seventh International Conference on Software Testing, Verification and Validation (ICST)},
  pages={153--162},
  year={2014},
  organization={IEEE}
}

@article{papadakis2015mutation,
  title={Metallaxis-FL: Mutation-based fault localization},
  author={Papadakis, Mike and Le Traon, Yves},
  journal={Software Testing, Verification and Reliability},
  volume={25},
  number={5-7},
  pages={605--628},
  year={2015},
  publisher={Wiley}
}

@inproceedings{wang2015blugram,
  title={Amalgam+: Composing rich information sources for accurate bug localization},
  author={Wang, Shaowei and Lo, David},
  booktitle={Journal of Software: Evolution and Process},
  volume={27},
  number={12},
  pages={921--942},
  year={2015},
  publisher={Wiley}
}

@inproceedings{li2019deepfl,
  title={DeepFL: Integrating multiple fault diagnosis dimensions for deep fault localization},
  author={Li, Xia and Li, Wei and Zhang, Yuqun and Zhang, Lingming},
  booktitle={Proceedings of the 28th ACM SIGSOFT International Symposium on Software Testing and Analysis (ISSTA)},
  pages={169--180},
  year={2019},
  organization={ACM}
}

@inproceedings{lou2021boosting,
  title={Boosting coverage-based fault localization via graph-based representation learning},
  author={Lou, Yiling and Zhu, Qihao and Dong, Jinhao and Li, Xia and Sun, Zeyu and Hao, Dan and Zhang, Lu and Zhang, Lingming},
  booktitle={Proceedings of the 29th ACM Joint European Software Engineering Conference and Symposium on the Foundations of Software Engineering (ESEC/FSE)},
  pages={664--676},
  year={2021},
  organization={ACM}
}

@article{ouyang2025repograph,
  title   = {{RepoGraph}: Enhancing AI Software Engineering with Repository-level Code Graph},
  author  = {Ouyang, Siru and Yun, Wenhao and Zeng, Yu and Yang, Zonghai and Zhang, Meng and Han, Jiawei},
  journal = {arXiv preprint arXiv:2410.14684},
  year    = {2025}
}

@article{chen2021codex,
  title     = {Evaluating Large Language Models Trained on Code},
  author    = {Chen, Mark and Tworek, Jerry and Jun, Heewoo and Yuan, Qiming and Pinto, Henrique Ponde de Oliveira and Kaplan, Jared and Edwards, Harri and Burda, Yuri and Joseph, Nicholas and Brockman, Greg and others},
  journal   = {arXiv preprint arXiv:2107.03374},
  year      = {2021}
}
\end{document}